\documentclass[
aps,
prd,
amsmath,amssymb,
reprint,
showpacs,
superscriptaddress,
longbibliography,
floatfix,
nobalancelastpage,
nofootinbib,
]{revtex4-2}

\usepackage{microtype}
\usepackage[usenames,svgnames,dvipsnames]{xcolor}
\usepackage{graphicx}
\usepackage{dcolumn}
\usepackage{bm}
\usepackage[
  colorlinks=true,
  linkcolor=MediumBlue,
  citecolor=MediumBlue,
  urlcolor=MediumBlue,
  ]{hyperref}
\usepackage{float}
\usepackage{comment}
\usepackage{orcidlink}
\usepackage{lineno}
\usepackage[T1]{fontenc}

\usepackage[
    multi-part-units=single,
    range-phrase=--,
    range-units=single,
    retain-explicit-plus=true,
    retain-unity-mantissa=false,
    separate-uncertainty=true,
    ]{siunitx}

\usepackage[capitalise]{cleveref}
\usepackage{soul}

\usepackage{xspace}

\newcommand{\sqzmaxlho}{5.2}
\newcommand{\sqzmaxllo}{6.1}

\newcommand{\maxLHOrange}{165}
\newcommand{\maxLLOrange}{177}
\newcommand{\medianLHOrange}{152}
\newcommand{\medianLLOrange}{160}
\newcommand{\LHOduty}{65.0}
\newcommand{\LLOduty}{71.2}
\newcommand{\coincDuty}{52.6}

\newcommand{\FP}{Fabry--P\'{e}rot\xspace}

\newcommand{\specificthanks}[1]{\@fnsymbol{#1}}

\begin{document}

\preprint{APS/123-QED}

\title{Advanced LIGO detector performance in the fourth observing run}


\author{E.~Capote\,\orcidlink{0009-0007-0246-713X}}
\thanks{These authors contributed equally to this work.}
\affiliation{Syracuse University, Syracuse, NY 13244, USA}
\affiliation{LIGO Hanford Observatory, Richland, WA 99352, USA}
\affiliation{LIGO Laboratory, California Institute of Technology, Pasadena, CA 91125, USA}

\thanks{ecapote@caltech.edu}

\author{\hspace{-0.15cm}$^{, \hspace{0.05cm}\dagger}$\ W.~Jia\,\orcidlink{0000-0002-5119-6328}}
\thanks{These authors contributed equally to this work.}
\affiliation{LIGO Laboratory, Massachusetts Institute of Technology, Cambridge, MA 02139, USA}

\author{N.~Aritomi\,\orcidlink{0000-0003-4424-7657}}
\thanks{These authors contributed equally to this work.}
\affiliation{LIGO Hanford Observatory, Richland, WA 99352, USA}

\author{M.~Nakano\,\orcidlink{0000-0001-7703-0169}}
\thanks{These authors contributed equally to this work.}
\affiliation{LIGO Laboratory, California Institute of Technology, Pasadena, CA 91125, USA}
\affiliation{LIGO Livingston Observatory, Livingston, LA 70754, USA}

\author{V.~Xu\,\orcidlink{0000-0002-3020-3293}}
\thanks{These authors contributed equally to this work.}
\affiliation{LIGO Laboratory, Massachusetts Institute of Technology, Cambridge, MA 02139, USA}
\affiliation{University of California, Berkeley, CA 94720, USA}

\author{\\R.~Abbott}
\affiliation{LIGO Laboratory, California Institute of Technology, Pasadena, CA 91125, USA}
\author{I.~Abouelfettouh}
\affiliation{LIGO Hanford Observatory, Richland, WA 99352, USA}
\author{R.~X.~Adhikari\,\orcidlink{0000-0002-5731-5076}}
\affiliation{LIGO Laboratory, California Institute of Technology, Pasadena, CA 91125, USA}
\author{A.~Ananyeva}
\affiliation{LIGO Laboratory, California Institute of Technology, Pasadena, CA 91125, USA}
\author{S.~Appert}
\affiliation{LIGO Laboratory, California Institute of Technology, Pasadena, CA 91125, USA}
% ...

\date{\today}

\begin{abstract}
On May 24th, 2023, the Advanced Laser Interferometer Gravitational-Wave Observatory (LIGO), 
joined by the Advanced Virgo and KAGRA detectors, began the fourth observing run for a two-year-long dedicated search for
gravitational waves. The LIGO Hanford and Livingston detectors have achieved an unprecedented sensitivity to gravitational waves, with an angle-averaged median range to binary neutron star mergers of \qty{\medianLHOrange}{Mpc} and \qty{\medianLLOrange}{Mpc}, and duty cycles of \LHOduty\% and \LLOduty\%, respectively, with a coincident duty cycle of \coincDuty\%. The maximum range achieved by the LIGO Hanford detector is \qty{\maxLHOrange}{Mpc} and the LIGO Livingston detector \qty{\maxLLOrange}{Mpc}, both achieved during the second part of the fourth observing run.
For the fourth run, the quantum-limited sensitivity of the detectors was increased significantly due to the higher intracavity power from laser system upgrades and replacement of core optics, and from the addition of a 300 m filter cavity to provide the squeezed light with a frequency-dependent squeezing angle, part of the A+ upgrade program. 
Altogether, the A+ upgrades led to reduced detector-wide losses for the squeezed vacuum states of light which, alongside the filter cavity, enabled broadband quantum noise reduction of up to \qty{\sqzmaxlho}{dB} at the Hanford observatory and \qty{\sqzmaxllo}{dB} at the Livingston observatory. 
Improvements to sensors and actuators as well as significant controls commissioning increased low frequency sensitivity. This paper details these instrumental upgrades, analyzes the noise sources that limit detector sensitivity, and describes the commissioning challenges of the fourth observing run. 
\end{abstract}

                              
\maketitle

\tableofcontents

\section{Introduction}\label{sec:intro}
Since the first detection of gravitational waves by the Advanced LIGO detectors at Hanford, Washington, and Livingston, Louisiana~\cite{LIGO} in September 2015~\cite{GW150914}, the field of gravitational-wave astronomy has rapidly advanced. In the subsequent observing run (O1), LIGO detected gravitational waves from another binary black hole merger~\cite{GW151226} and fully opened this new window of astronomy. During the second observing run (O2)~\cite{GWTC-1}, the Virgo detector in Cascina, Italy~\cite{Virgo}, joined the effort and the three-detector network made the first observation of gravitational waves from a binary neutron star merger ~\cite{GW170817-1, GW170817-3}, which was associated with electromagnetic counterparts~\cite{GW170817-2}. 
During the third observing run (O3), LIGO and Virgo identified 79 high-significance event candidates~\cite{O3a,O3b}. These detections included observations such as the first black hole merger with clear evidence for unequal masses~\cite{GW190412}, the first detection of black holes plausibly residing in the pair-instability mass-gap~\cite{GW190521}, and the first unambiguous discovery of neutron star-black hole binaries~\cite{NSBHO3}. These detections, among many others in the catalog, provide new perspectives on stellar evolution.

\begin{figure}[t]
    \centering
    \includegraphics[width=0.75\columnwidth, trim={4cm 0 4cm 0}]{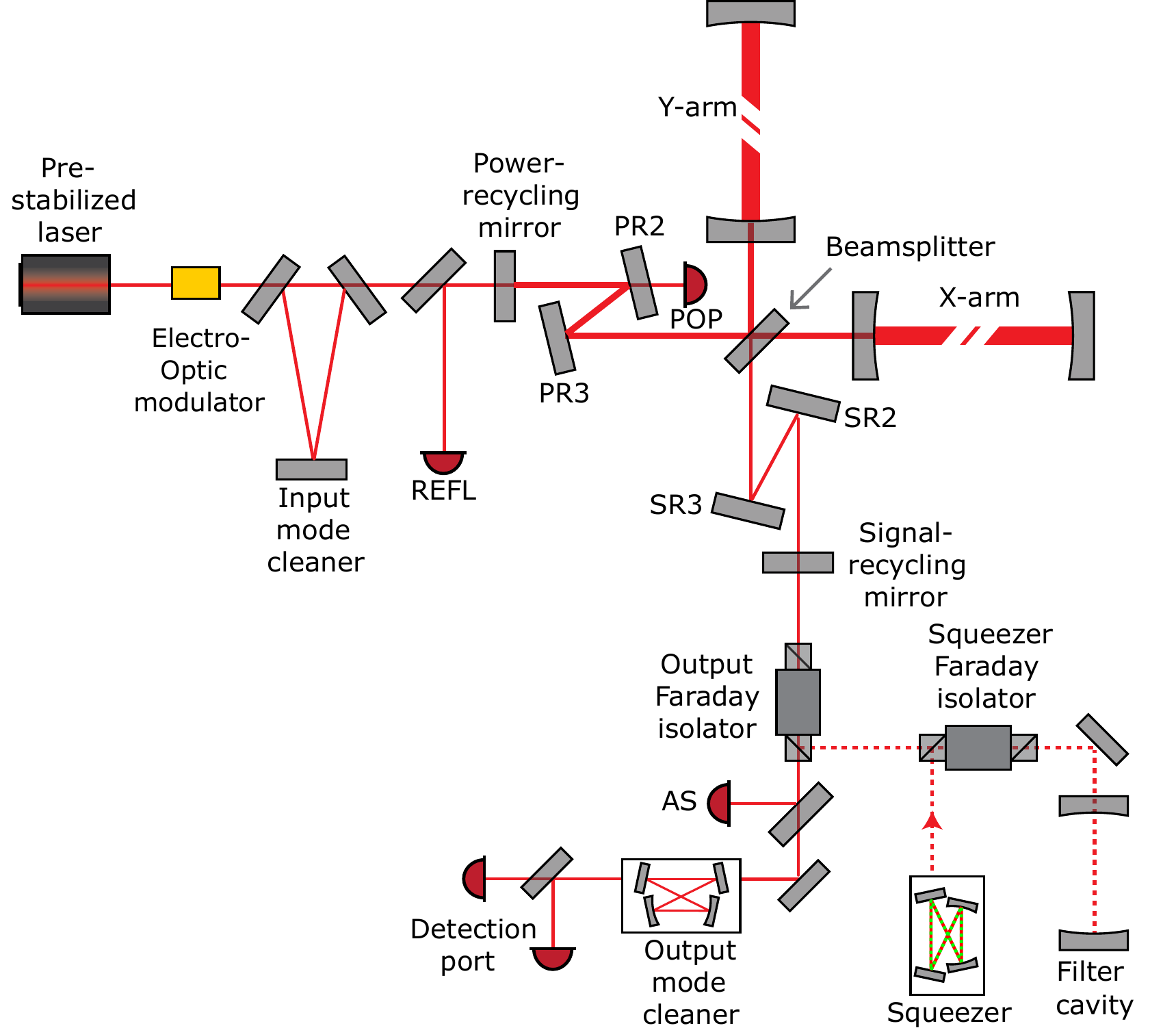}
    \caption{Schematic of the Advanced LIGO interferometer layout. The interferometer has two \qty{4}{\km} arm cavities and two recycling cavities. Each recycling cavity is folded with two mirrors. The input mode cleaner and the output mode cleaner cavities provide spatial mode filtering of the input and output beam of the interferometer. Squeezed vacuum is generated in the squeezer cavity
    , rotated upon reflection from the new 300 m filter cavity to acquire a frequency-dependent squeezing angle
    , and injected at the anti-symmetric port of the interferometer via the output Faraday isolator. The gravitational-wave signal is detected by two photodetectors in transmission of the output mode cleaner cavity. The electro-optic modulator before the input mode cleaner introduces phase modulation at three frequencies: 9 MHz, 45 MHz, and 118 MHz. These modulations are used for sensing and control of the auxiliary degrees of freedom using photodetectors and wavefront sensors at the interferometer reflection port (REFL), at the pick-off of the power recycling cavity port (POP), and at the anti-symmetric port (AS) after the output Faraday isolator. A full description of the detector operation can be found in~\cite{o3paper}, and that of the new filter cavity subsystem in~\cite{ganapathyBroadbandQuantumEnhancement2023}.
    }
    \label{fig:ligo}
\end{figure}

After the completion of O3, both LIGO detectors were taken offline for a series of planned upgrades and maintenance.
These upgrades included the first part of the planned A+ upgrade program designed to further boost the sensitivity and performance of the advanced detectors~\cite{abbottProspectsObservingLocalizing2020}.
\cref{fig:ligo} shows the layout of the Advanced LIGO detectors in the fourth observing run (O4), following these instrumental upgrades.
The dual-recycled \FP{} Michelson detector topology and operation remained the same as the previous observing run~\cite{o3paper}, with the notable addition of a \SI{300}{m} filter cavity for frequency-dependent squeezing~\cite{ganapathyBroadbandQuantumEnhancement2023}.
The addition of this new squeezing system included extensive vacuum and facility construction, new suspended mirrors, and upgraded seismic isolation, electronics, and output optical components.
Further upgrades included replacing core interferometer optics, upgrading the pre-stabilized laser system, and improvements to the controls system.

O4 is now ongoing, comprising two separate parts from May 24, 2023 to January 16, 2024 (O4a) and from April 10, 2024 to June 9, 2025 (O4b). 
The Virgo detector joined the observation for O4b. The KAGRA detector in Japan~\cite{KAGRA} joined for the first month of O4a and is scheduled to join for the last few months of O4b. During O4a, the LIGO--Virgo--KAGRA collaboration announced 81 significant event candidates, including the merger of a neutron star with an unknown compact object in the lower mass gap, having an estimated mass of 3.6\(M_\odot\)~\cite{GW230529}, further enriching the understanding of compact-objects populations. 
The detector improvements summarized in this paper led to a marked increase in sensitivity;
the number of candidate events detected in O4 so far has reached 149, already greater than the 90 total events detected from O1 to O3, indicating the success of instrumental upgrades carried out between O3 and O4.

This work is organized as follows: 
\cref{sec:o4overview} summarizes the overall performance of the detectors, including astrophysical range, detection rate, and duty cycle.
\cref{sec:noise} presents the detector noise budgets for O4, and evaluation of the noise sources that limit detector sensitivity. 
\cref{sec:improvements} describes the instrumental upgrades made for O4. 
\cref{sec:characterization} discusses quantum noise investigations pursued to understand the quantum limits to total sensitivity, analysing both the squeezing performance and interferometer operating power. 
\cref{sec:highpower} summarizes the challenges faced in pursuing higher power operation for O4.
\cref{sec:future} reviews the plans for future detector improvements in the context of the successes and challenges of the current run.

\section{O4 Overview}\label{sec:o4overview}

The Advanced LIGO detectors achieved unprecedented sensitivity in the fourth observing run. This section details the performance and operation of the detectors in O4, including astrophysical range, detection rates, lock acquisition, and duty cycle. 

\subsection{Astrophysical range}\label{sec:range}
\begin{figure}[tbp]
    \centering
        \includegraphics[width=\linewidth]{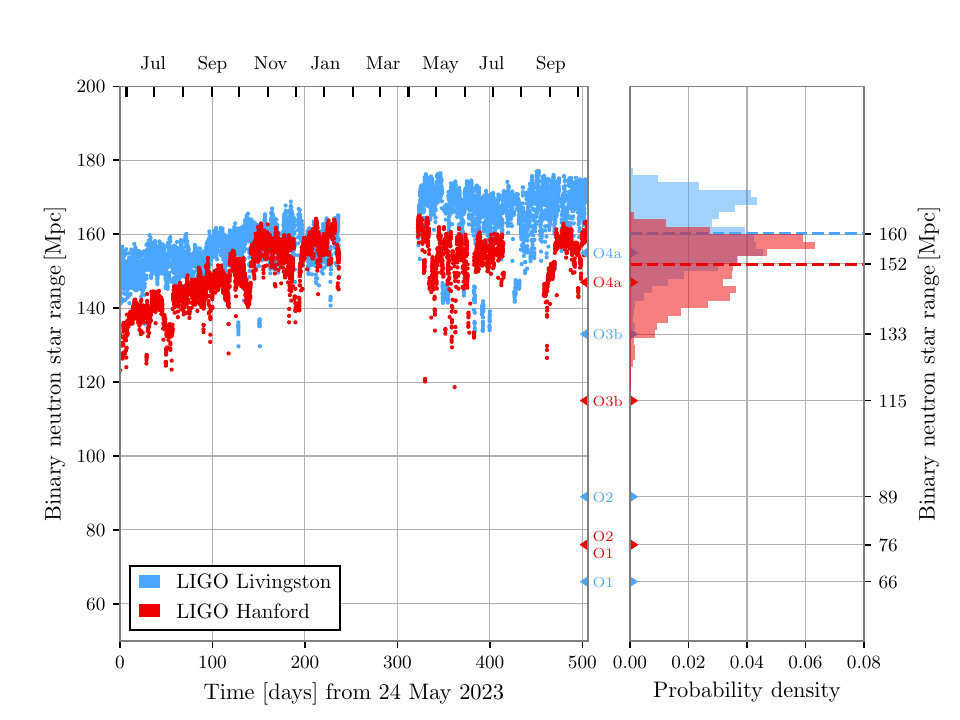}
    \caption{Astrophysical range trends for the LIGO Hanford (red) and LIGO Livingston (blue) observatories up to October 11, 2024, including the entirety of O4a and O4b*. The left plot shows the median hourly binary neutron star (BNS) range of the LIGO Hanford and LIGO Livingston detectors, whereas the right plot shows the histogram of the ranges. Both detectors have improved sensitivity compared to the median sensitivity in observing runs 1--3~\cite{GWTC-1,GWTC-2,O3a,O3b}. Both detectors increased sensitivity over the course of O4, especially the Hanford detector. During O4a, the detectors achieved median BNS ranges of \qty{148}{Mpc} (Hanford) and \qty{156}{Mpc} (Livingston). Improvements during the break between O4a and O4b (data gap) increased the detector median ranges to \qty{\medianLHOrange}{Mpc} (Hanford) and \qty{\medianLLOrange}{Mpc} (Livingston), with a maximum hourly-median range of \qty{\maxLHOrange}{Mpc} (Hanford) and \qty{\maxLLOrange}{Mpc} (Livingston) achieved during O4b*. The Hanford detector spent additional time offline during O4b* for emergency repairs to the output optics, indicated by the second gap in the Hanford detector data (discussed in \cref{sec:dutyCycle}).}
    \label{fig:rangeplots}
\end{figure}

\cref{fig:rangeplots} shows the binary neutron star (BNS) range of the LIGO detectors during O4, covering the entire O4a run and the first six months of O4b (referred to here as O4b*, from April 10, 2024 to October 11, 2024). 
The BNS range provides a standard metric of sensitivity based on the calibrated strain sensitivity of the detectors. It is the distance to which a $1.4\ M_{\odot}$--$1.4\ M_{\odot}$ BNS merger can be observed with an amplitude signal-to-noise ratio (SNR) of 8, accounting for the antenna pattern of the two LIGO detectors, as described in~\cite{Chen:2017wpg}. For O4, the strain calibration uncertainty was measured to be 10\% in magnitude and $10^\circ$ in phase from \SI{20}{\Hz} to \SI{2}{\kHz}~\cite{O4cal}.

In O4, the detectors achieved a BNS range of \qtyrange{130}{165}{Mpc} for LIGO Hanford and \qtyrange{145}{177}{Mpc} for LIGO Livingston. Detector commissioning work continued during O4, increasing the detector sensitivity at both sites. \cref{fig:rangeplots} shows that the O4 median range for the Livingston detector is \qty{\medianLLOrange}{Mpc}, while the Hanford detector has a median range of \qty{\medianLHOrange}{Mpc}. 
The significant increases in the Hanford detector sensitivity during the run came from the reduction of input power shortly into the run that enabled further low-frequency noise improvements (June 2023, described in \cref{sec:powerreduction}), and the recovery of squeezer crystal losses which allowed squeezed light to further reduce detector quantum noise (October 2023, described in \cref{sec:sqz_level}). A significant increase in the Livingston detector sensitivity came 
from cleaning the end test masses during the break between O4a and O4b (described in \cref{sec:testmasses}). 
An additional observing break was taken at the Hanford observatory in O4b for emergency repairs, further described in \cref{sec:dutyCycle}. This break is evident in the lack of range data from the Hanford observatory between July and August in \cref{fig:rangeplots}.

\begin{figure}[tbp]
    \centering
    \includegraphics[width=0.48\textwidth, trim={0mm 0 0 0}]{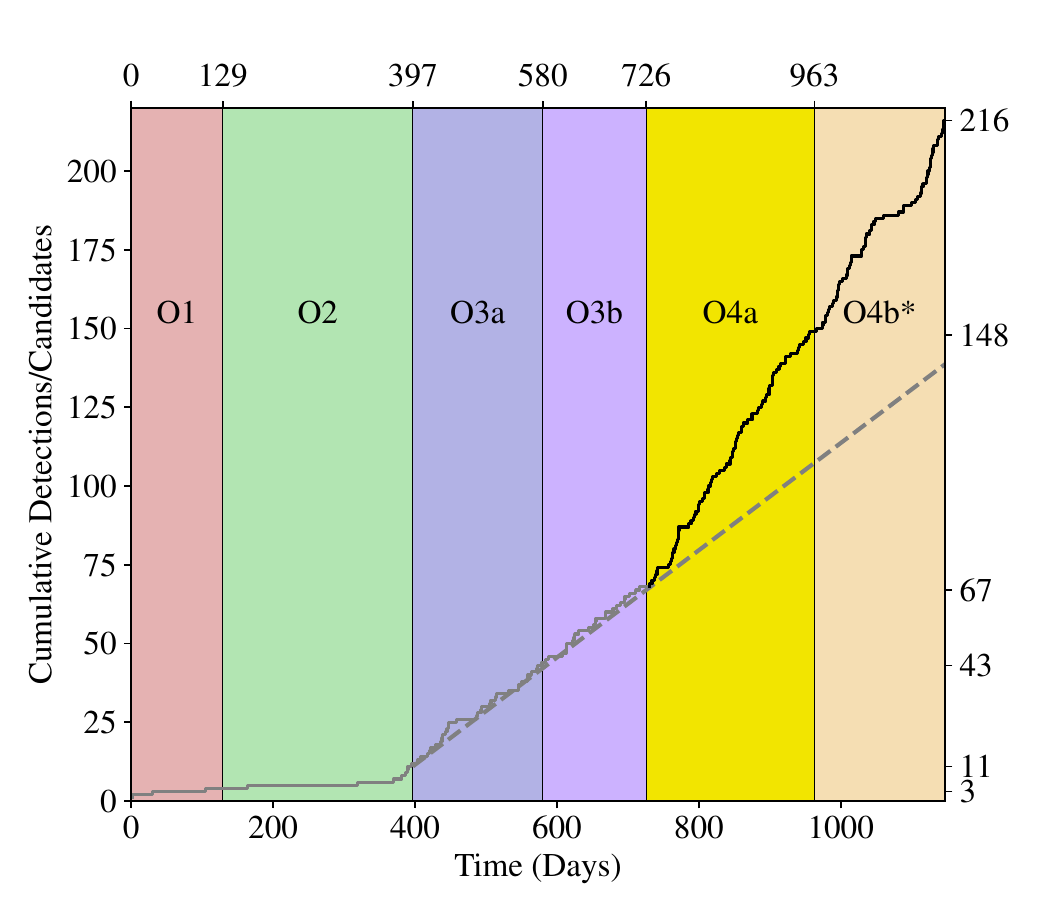}
    \caption{Cumulative detections and candidates up to October 11, 2024, including the entirety of O4a and O4b*. Events represented in O1 and O2 are all the identified candidates from that observing period, as announced in GWTC-1~\cite{GWTC-1}. Candidates in O3 and O4 only include preliminary candidates identified in online triggers~\cite{O3b,gracedb}. Including online and offline detections, a total of 90 events were observed prior to O4~\cite{GWTC-1,O3a,O3b}. In O4a, 81 candidates were identified in online triggers. The grey dotted line represents the event detection rate from O3, projected into the O4 observing time. The difference between the grey dotted line and the black event detection line in O4 highlights how the advancement in detector sensitivity results in the increased event detection rate of the network. The small plateau in the event rate during O4b* corresponds to the time the Hanford observatory was down for emergency repairs, discussed in \cref{sec:dutyCycle}.
    }
    \label{fig:o4_event_rate}
\end{figure}

At O4 sensitivities, the event detection rate is approximately doubled relative to O3. \cref{fig:o4_event_rate} shows the cumulative number of events as a function of the total observing time in days. 
During O4a, a total of 81 non-retracted, high-significance public alerts were released after 237 days of observing~\cite{gracedb}. By comparison, a total of 56 online triggers were registered in O3 after 329 days of observing~\cite{gracedb,GWTC-2,O3a,O3b}. These alerts are event candidates that generate an online trigger in the low-latency pipeline. Online triggers are vetted to confirm that data quality issues do not impact the detection or analysis of the signals~\cite{Soni:2024isj}.

In the 726 days of observing over the first three runs, a total of 90 gravitational-wave events were detected~\cite{GWTC-1,GWTC-2,O3a,O3b}. For O3, the total number of detections (79) included candidates found in online low-latency triggers (56) and in offline searches after the run. Offline searches for gravitational-wave event candidates within the O4a data are ongoing, but it is expected that the total number of gravitational-wave event candidates will increase beyond the online candidates, as in O3. As of October 11, 2024, an additional 68 non-retracted, high-significance public alerts were released during the first six months of O4b.

\cref{fig:o4_event_rate} compares the event rate during O4a and O4b*, with that of the previous observing runs~\cite{GWTC-1,GWTC-2,O3a,O3b,gracedb}. The event rate shown for O3 only includes non-retracted online triggers~\cite{O3b,gracedb}, and does not include additional events found offline. The event rate for O1 and O2 represents the total number of events detected in those runs, as online triggers were not implemented at the time~\cite{GWTC-1}. A comparison of the online event trigger rate in O3 to the online trigger rate in O4 demonstrates the significant improvement in sensitivity to gravitational-wave events achieved through the upgrades performed between O3 and O4.

\begin{figure}[t]
    \centering
    \includegraphics[width=\linewidth,trim={0 0 0 0}]{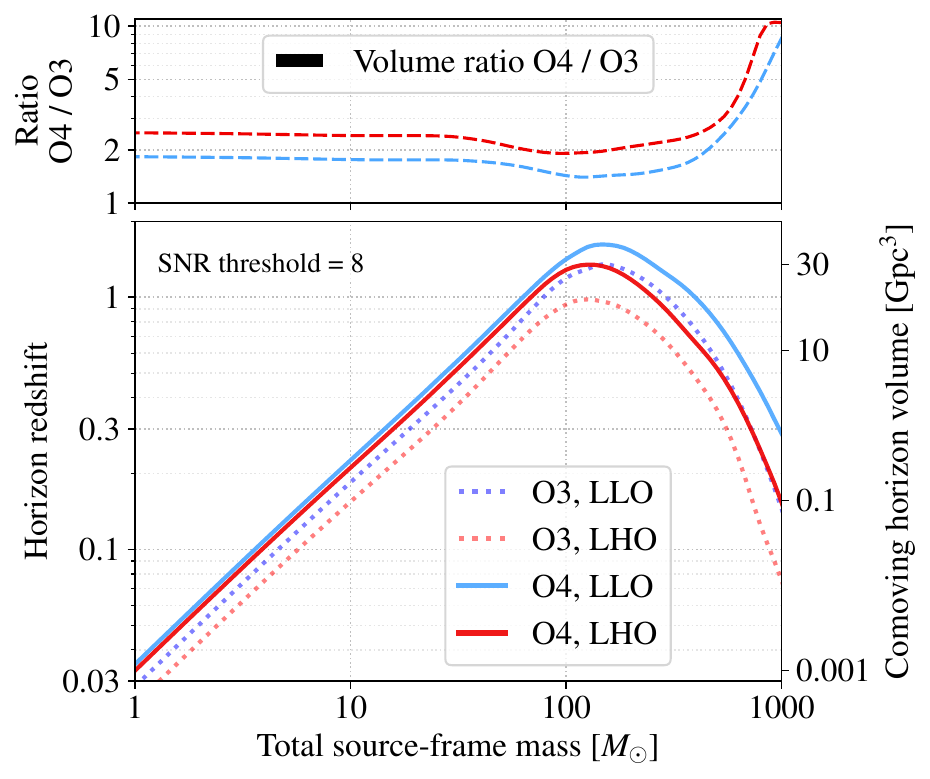}
    \caption{
    Observable horizons shown as the horizon redshift (lower plot, left axis) and co-moving horizon volume (lower plot, right axis)~\cite{Chen:2017wpg} for the detector noise in O4a (\cref{fig:noisebudgets}). Compared to the O3 sensitivities~\cite{o3bH1strain,o3bL1strain}, instrument upgrades increased the observable volume of the detectors by up to 2.5-fold (Hanford) and 1.8-fold (Livingston) for compact object mergers with total mass below $100M_\odot$.  
    The lower plot shows the horizons for an event detection with signal-to-noise ratio (SNR) of 8 as a function of the total source-frame mass. The upper plot shows the volume ratio of O4a compared to O3.}
    \label{fig:horizonplots}
\end{figure}

The detector BNS range provides a broad metric to characterize detector performance. However, detector improvements, especially at low frequency, have increased the sensitivity to a variety of astrophysical sources. \cref{fig:horizonplots} plots the increased observable horizons as a function of the total source-frame mass, showing an increase in sensitivity to more distant events (i.e., mergers at higher redshifts) and heavier compact object mergers, characterized using the methodology described in~\cite{Chen:2017wpg}. 
Generally, detector improvements have lead to an increase in signal-to-noise ratio from O3 to O4 across a wide range of compact object merger masses, with the most significant improvement ratio at high total source-frame mass (low frequency).

\subsection{Lock acquisition}

In order to detect gravitational-wave signals, the interferometers must be in the ``locked'' state: all mirror lengths and alignments are controlled within a linear regime, and light stably resonates inside the optical cavities~\cite{Staley_2014}. The procedure to bring the interferometer from an uncontrolled state to this stable state is called ``lock acquisition,'' which is described in~\cite{O1paper, O1ligo, o3paper}. 

Because the automated detector locking process has been further streamlined from O3~\cite{o3paper}, the higher operating powers and additional subsystems have not significantly changed the lock acquisition time. 
A primary difference in the lock acquisition process compared to O3~\cite{o3paper} is the addition of the filter cavity to prepare the squeezed vacuum states of light with a frequency-dependent squeeze angle. The squeezer and filter cavity are locked in parallel with the main interferometer, with locking procedures detailed previously in~\cite{ganapathyBroadbandQuantumEnhancement2023}. 
Once the main interferometer is locked, the filter cavity is locked, and frequency-dependent squeezed light is injected. Shortly after, the interferometer enters observation mode.

If a disturbance causes a controlled degree of freedom to move outside the linear control regime, a ``lock loss'' occurs. The locking process must then begin again. 
Lock losses can be caused by a variety of phenomena, such as earthquakes, storms, high winds, control instabilities, and instrumental glitches. Still, the majority of locklosses occur for unknown reasons.

\subsection{Duty cycle}\label{sec:dutyCycle}

The duty cycles for observing in O4, including all of O4a and O4b*, are \LHOduty\% for Hanford and \LLOduty\% for Livingston; the duty cycle for observing in dual coincidence with both LIGO detectors is \coincDuty\%.
Time spent out of observing can be divided into planned commissioning time where detector improvements are made, scheduled maintenance, lock acquisition time, or time spent unlocked due to unfavorable environmental conditions such as earthquakes, elevated microseismic activities, storms, or high winds, similar to O3~\cite{o3paper}. 

The greatest impacts to the Hanford observatory duty cycle included control instabilities due to high power operation at the start of the run (see \cref{sec:highpower} for more details) and damage to the output optics that required emergency intervention during O4b. The Livingston observatory duty cycle was impacted early in O4a due to logging activities near the detector.

After the start of O4b, a lockloss at the Hanford observatory caused damage to one of the optics in the output Faraday isolator (see \cref{fig:ligo}). By adjusting the output alignment to avoid the damaged area on the optic, observing continued for a few months with only minimal effect on detector performance. After another lockloss caused more damage on July 12, 2024, lock acquisition and observing were no longer possible. Therefore, shortly after, the Hanford observatory halted observing to perform an emergency incursion into the vacuum chambers that house the output optics. Two damaged optics within the output Faraday isolator were replaced. On August 24, 2024, the Hanford observatory rejoined the observing run. This gap in observing data is evident in \cref{fig:rangeplots}, and in the small plateau in event detection in \cref{fig:o4_event_rate}.

On October 11, 2024, a hardware failure in the fast shutter between the output Faraday isolator and the output mode cleaner (see \cref{fig:ligo}) at the Livingston observatory occurred. After this date, Livingston briefly ceased observing for an emergency incursion to repair this equipment.

\section{Instrument noise}\label{sec:noise}

The LIGO Hanford and Livingston detector sensitivities are limited by various fundamental and technical noises. Fundamental noises often cannot be improved without altering the basic parameters of the interferometer itself, such as laser power, mirror mass, squeezing injection, or material properties. On the other hand, technical noises can be affected by improvements to the hardware and software associated with the interferometer operation, including sensors, actuators, control filters, circuits, and dampers that mitigate scattered light.

\begin{figure*}[htbp]
    \centering
    \begin{minipage}{0.85\textwidth}
        \centering
        \includegraphics[width=\linewidth]{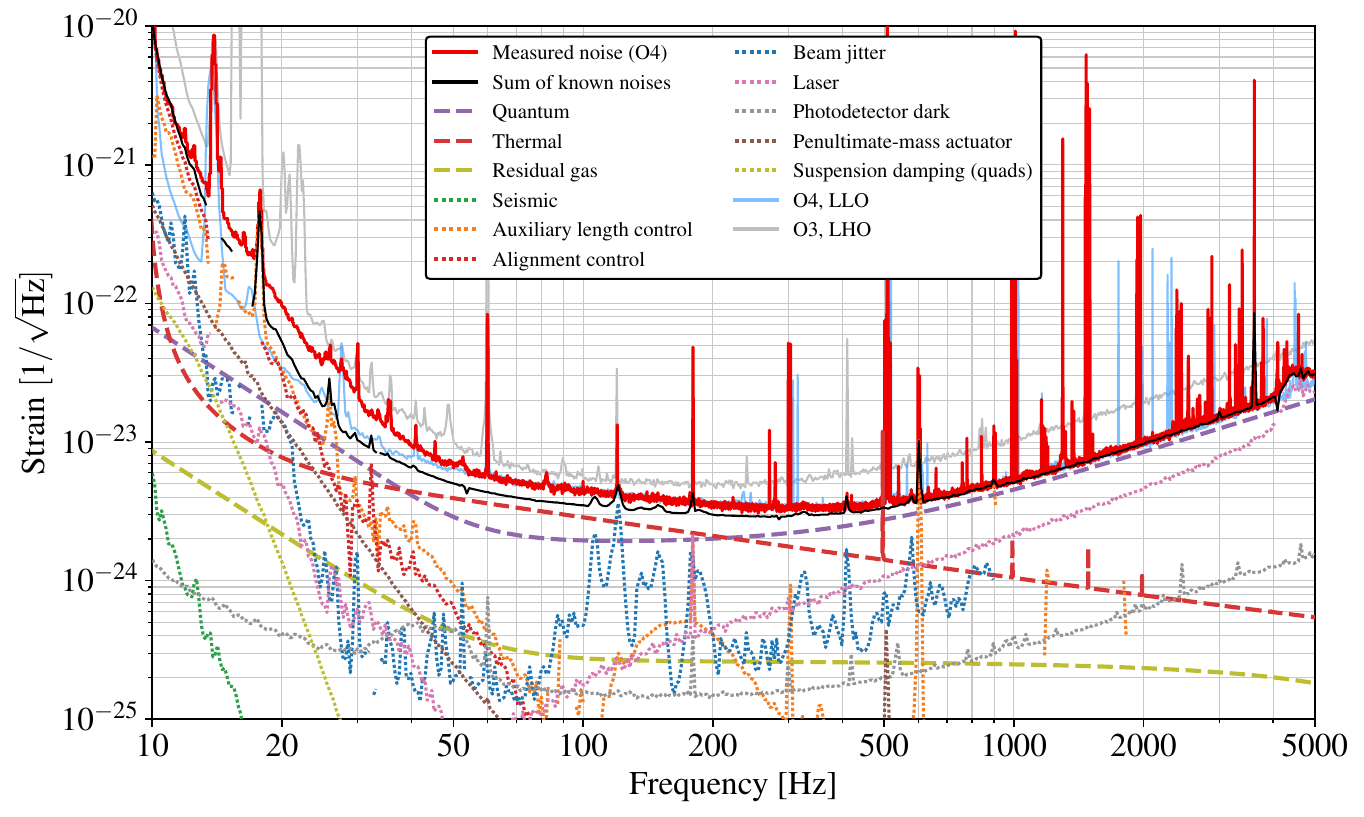}
        \par{(a) Noise budget for the LIGO Hanford Observatory, as of December 2023.} 
        \label{fig:LHO_NB}
    \end{minipage}
    \begin{minipage}{0.85\textwidth}
        \centering
        \includegraphics[width=\linewidth]{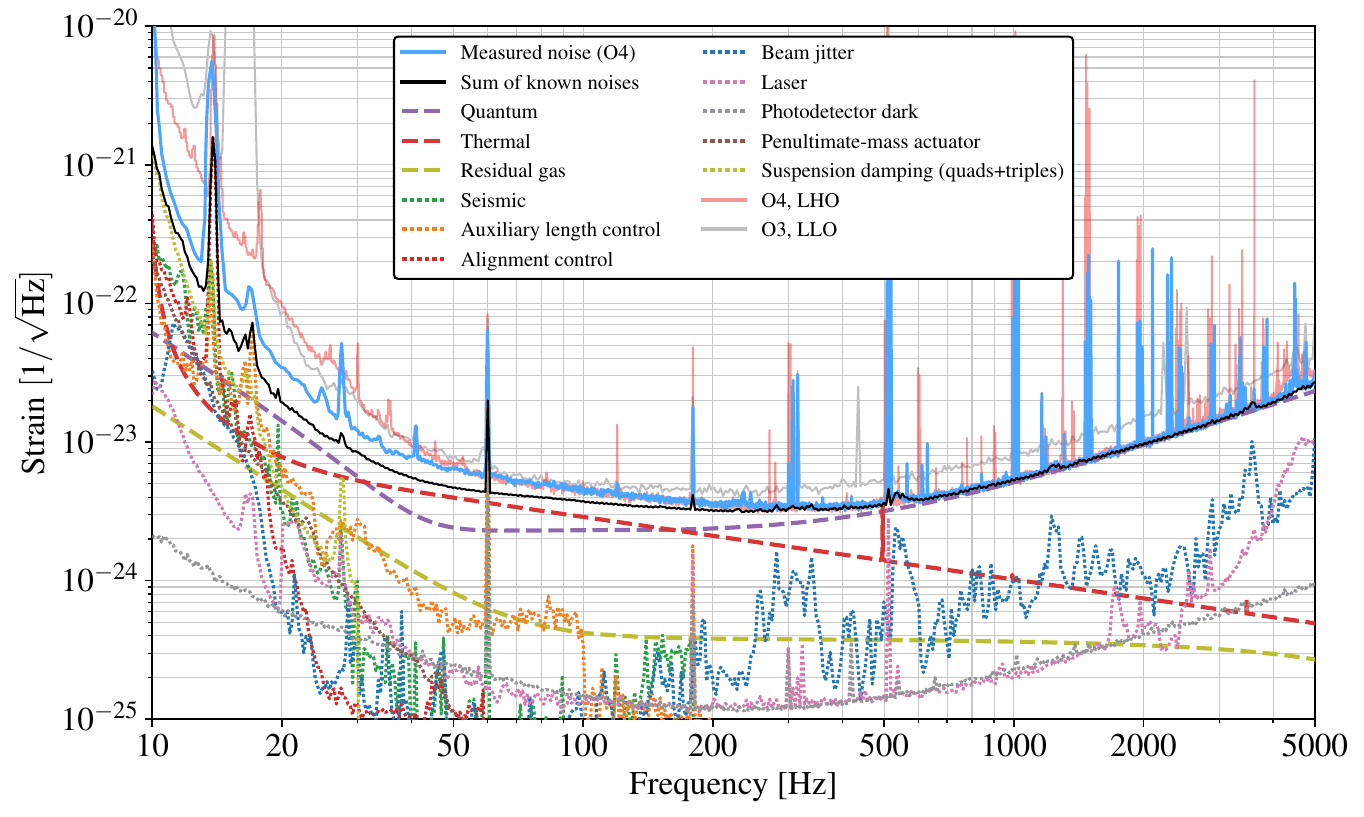}
        \par{(b) Noise budget for the LIGO Livingston Observatory, as of October 2023.} 
        \label{fig:LLO_NB}
    \end{minipage}
    \caption{Noise budgets for the LIGO Hanford Observatory (upper) and LIGO Livingston Observatory (lower) in strain noise amplitude spectral density units. Dashed lines show noises calculated from theoretical understanding of detector limitations or ex-situ measurements of detector components. Dotted lines represent noises that are measured and projected from auxiliary witness channels. Red (blue) lines show the total measured noise at Hanford (Livingston), while black lines show the expected total noise (sum of known noises). Both detectors face similar limiting noises: controls noise from angular, length, and local suspension damping below \qty{50}{\Hz}, a mix of quantum noise and thermal noise between \qtyrange{50}{500}{\Hz}, and shot noise above 500 Hz. Discrepancies between the measured and expected noise are greatest at low frequencies. All spectra shown use mean averaging for power spectral density estimation. Individual noise terms are discussed in \cref{sec:noise}.
    }
    \label{fig:noisebudgets}
\end{figure*}

Budgets of the limiting noise sources for the LIGO detectors in O4a are shown in \cref{fig:noisebudgets}. 
Dashed traces in \cref{fig:noisebudgets} represent noises that are calculated from detector properties or ex-situ measurements of detector components. Dotted traces in \cref{fig:noisebudgets} represent noises whose coupling is measured and then directly projected from auxiliary witness channels, as described in Refs.~\cite{O1paper,o3paper}. The sum of these expected noises (black) is compared to the total measured strain noise at each detector. Discrepancies between the expected and measured noise indicate areas where the limiting noise is not well-understood, or where the noise is nonstationary. The sensitivity from the previous observing run (O3) is shown for comparison~\cite{o3bH1strain,o3bL1strain}; \cref{fig:noisebudgets} shows both detectors have improved sensitivity across the detection band in O4 relative to O3.

At frequencies above \qty{50}{\Hz}, the two LIGO detectors face similar limiting noises.
At kilohertz frequencies, the two detectors achieve comparable shot-noise-limited strain sensitivities, with slight differences due to the different circulating arm powers and squeezing efficiencies.
Between roughly \qtyrange{50}{500}{\Hz}, detector sensitivity is primarily limited by thermal noise, and secondarily by quantum noise, with notable discrepancies between measured noise and expected total noise. 
Below \qty{50}{\Hz}, the differences between the Hanford and Livingston detectors are most apparent. This is largely attributable to differences in controls noise, seismic noise, and input beam jitter. Both detectors also have some amount of unknown noise, which could be due to scattered light, which is not estimated here, or from nonstationary noise contributions to the total noise~\cite{Soni:2024isj, Soni_2021, LIGO:2021ppb, scattershelf}. Throughout the detection band, both detectors are limited by narrow instrumental artifacts from sources such as mechanical modes of suspensions, electromagnetic couplings, or unknown environmental couplings~\cite{Soni:2024isj}. Most of these narrow features are not budgeted in \cref{fig:noisebudgets}, but many are evident in the total measured noise of each detector.

Of the noises shown in \cref{fig:noisebudgets}, the residual gas and seismic noise estimates are the same as in O3~\cite{o3paper}. Residual gas noise is estimated through the partial pressures of the various gas species present in the ultrahigh vacuum systems of each detector~\cite{o3paper}. Seismic noise is suppressed in the detector through multi-stage pendula, suspended seismic isolation platforms, and a hydraulic pre-isolation system~\cite{Aston2012,Daw_2004,Matichard_2015,HEPI_2014}. The residual seismic noise shown in \cref{fig:noisebudgets} is estimated through the measurement of residual mirror motion that is linearly coupled to the gravitational-wave strain~\cite{o3paper}. 
This section describes the detector noise sources in O4 (\cref{fig:noisebudgets}) that differ from O3~\cite{o3paper}, and presents an updated understanding of the noises that have limited performance of the LIGO detectors in O4.

\subsection{Quantum noise} \label{sec:sqz}

Quantum noise limits detector sensitivity across most of the astrophysical band. Quantum noise originates from vacuum fluctuations entering the detector's antisymmetric port~\cite{Caves1981}, where the two quadratures of the optical vacuum field manifest as two forms of quantum noise: shot noise and quantum radiation pressure noise. Shot noise arises from fluctuations in the phase quadrature of the vacuum field entering the antisymmetric port, and it is the dominant noise source above hundreds of hertz. The signal-to-shot-noise ratio increases with the square root of the intracavity power. Quantum radiation pressure noise is a displacement noise that results from vacuum fluctuations in the amplitude quadrature, which induce a fluctuating force on the optics through radiation pressure. These fluctuations are attenuated by the mechanical response of the test mass mirror suspensions, and can become a limiting noise source below about 50 Hz. In power spectral density, quantum radiation pressure noise increases proportionally to the arm power. 
The laser system upgrades (\cref{sec:psl}) and optics upgrades (\cref{sec:testmasses}) allowed arm powers to be increased for O4; methods of estimating arm powers are described in \cref{sec:arm_power_characterization}. 

Starting in O3, the LIGO detectors also began to inject non-classical states of light, known as ``squeezed vacuum,'' at the anti-symmetric port to reduce quantum noise~\cite{Caves1981,wuGenerationSqueezedStates1986,vahlbruchObservationSqueezedLight2008,GEOgravitationalWaveObservatoryShotNoiseLimit2011} during astrophysical observing~\cite{tsePRL19QuantumEnhancedAdvanced,yuN20QuantumCorrelations,mccullerPRD21LIGOQuantum,barsottiSqueezedVacuumStates2018,galaxies10020046}.Squeezed vacuum uses quantum correlations between photon pairs to reduce the noise variance in one quadrature of the squeezed state at the expense of increased noise in the conjugate quadrature, e.g, reducing fluctuations in phase at the cost of increasing fluctuations in amplitude~\cite{Caves1981,schnabelSqueezedStatesLight2017a}. 
For O4, a filter cavity was added to the squeezing system to prepare frequency-dependent squeezed vacuum states of light for broadband quantum noise reduction~\cite{kimblePRD01ConversionConventional,ganapathyBroadbandQuantumEnhancement2023}. The squeezing system upgrades for O4 are summarized in \cref{sec:sqzUpgrades} and described in detail in Refs.~\cite{ganapathyBroadbandQuantumEnhancement2023,jiasubSQL}.

\cref{sec:characterization} describes the quantum noise analysis used to determine key parameters of the quantum noise models~\cite{bncPhysRevD.64.042006,mccullerPRD21LIGOQuantum}, including the squeezer parameters (Sec.~\ref{sec:sqz_characterization}) and the interferometer arm power (Sec.~\ref{sec:arm_power_characterization}). 
From these estimates of squeezer and interferometer parameters, the Gravitational-Wave Interferometer Noise Calculator (GWINC)~\cite{gwinc} is used to calculate the squeezed quantum noise models plotted in the dashed purple traces of \cref{fig:noisebudgets}.

\subsection{Thermal noise} \label{sec:ctn}

\begin{figure}
    \centering
    \includegraphics[width=\columnwidth, trim={0 0 0 0}
    ]{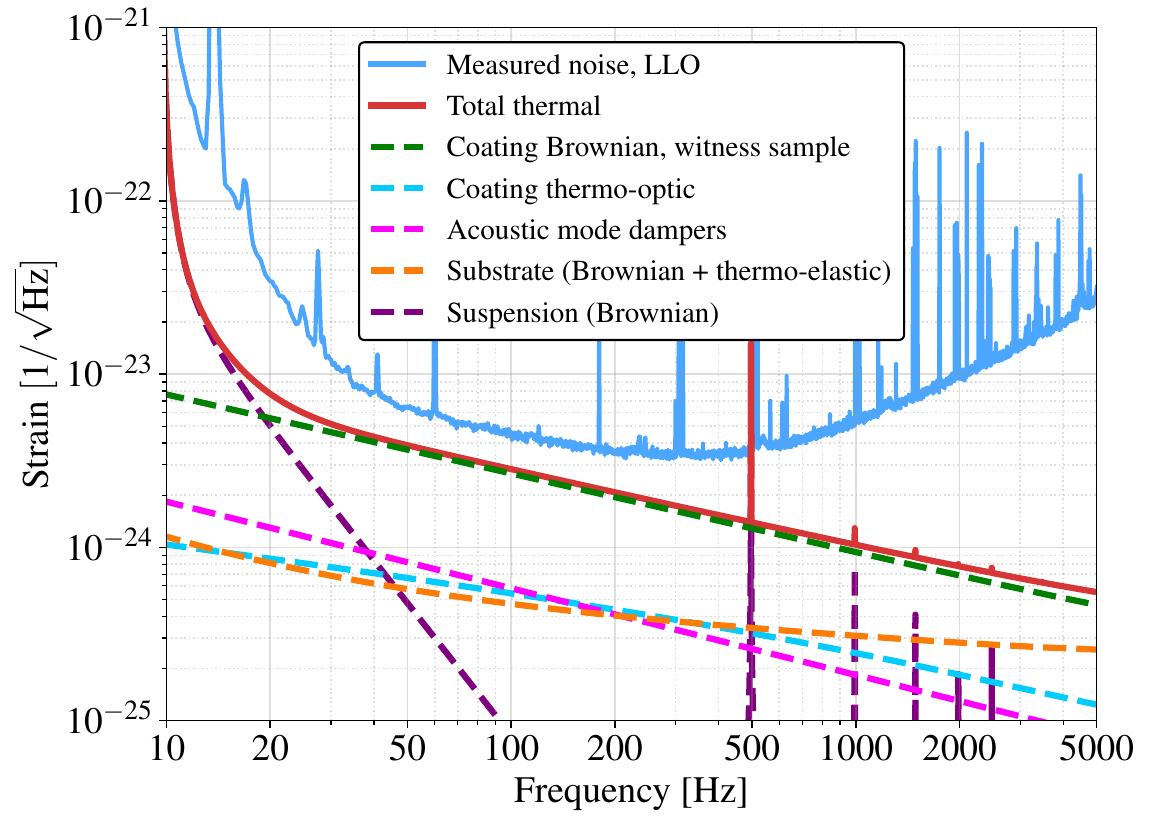}
    \caption{Thermal noise sub-budget compared to the total strain noise for the Livingston detector. 
    Coating Brownian noise is shown with the amplitude and frequency-dependence measured from witness samples of the Advanced LIGO test mass coating run (green dashed trace)~\cite{amatoOpticalMechanicalProperties2021}, and is expected to be the dominant source of thermal noise across most of the detection band.
    }
    \label{fig:llo_thermal_subbudget}
\end{figure}

Thermal motion in the mirror suspensions, mirror substrates, and optical coatings contributes noise in gravitational-wave detectors. The most significant noise source of these three in Advanced LIGO is coating thermal noise, arising from Brownian motion in the optical mirror coatings, limiting the detector sensitivity around 100 Hz~\cite{BRAGINSKYthermodynamical}. The Advanced LIGO test masses are mirrors using Bragg reflectors to create high reflectivity surfaces. The Bragg reflector consists of alternating layers of low- and high-index of refraction materials~\cite{multimaterial}. In the Advanced LIGO case, the low index material is silica, SiO$_2$, and the high index material is tantala doped with titania, TiO$_2$:Ta$_5$O$_2$~\cite{Harry_2007,Granata_2020}. Thermal noise arising from an incident Gaussian beam can be shown, via the fluctuation--dissipation theorem, to depend on the loss angle of the material~\cite{levinthermalnoise, CallenFDtheorem, saulsonThermal,Hong:2012jv}. Of those two materials, titania-doped tantala has a higher loss angle, and it is the loss from this material that sets the current coating thermal noise limit in Advanced LIGO.

To estimate the coating thermal noise in the Advanced LIGO interferometers, smaller witness samples were produced during the same test mass coating processes to allow for representative direct external measurements of coating thermal noise~\cite{grasDirectMeasurementCoating2018}.  
The estimated coating thermal noise level for these Advanced LIGO witness samples was reported in Ref.~\cite{amatoOpticalMechanicalProperties2021} as
\begin{equation}\label{eq:CTN}
    \sqrt{S_\text{CTN}} = 1.13\times10^{-20} \left(\frac{100~\text{Hz}}{f}\right) ^{0.45} \frac{\text{m}}{\sqrt{\text{Hz}}} .
\end{equation}
The total thermal noise trace in \cref{fig:noisebudgets} uses this estimate for coating thermal noise. 

\cref{fig:llo_thermal_subbudget} shows a sub-budget of the total thermal noise, compared to the total Livingston detector strain noise. 
Coating thermal noise is expected to be the dominant contribution to the total thermal noise above \qty{30}{Hz}, excluding the narrow suspension resonances. 
In order to suppress the contribution of substrate thermal noise to the total noise, the test mass substrates are made of high-quality fused silica to achieve a low mechanical loss. 
The resulting mechanical resonances of the substrates have a very high quality factor ($Q$) that can lead to the opto-mechanical parametric instabilities (PIs) described in \cref{sec:PImitigation}. Acoustic mode dampers (AMDs) were installed to reduce the $Q$ of these mechanical resonances~\cite{biscansAMDs}. Based on external measurements, the thermal noise from AMDs is estimated to contribute less than 1\% (in amplitude) to the total thermal noise~\cite{biscansAMDs}.

\subsection{Laser noise} \label{sec:lasernoise}
\begin{figure}[]
    \centering
    \includegraphics[width=0.48\textwidth, trim={0 0 0 0}]{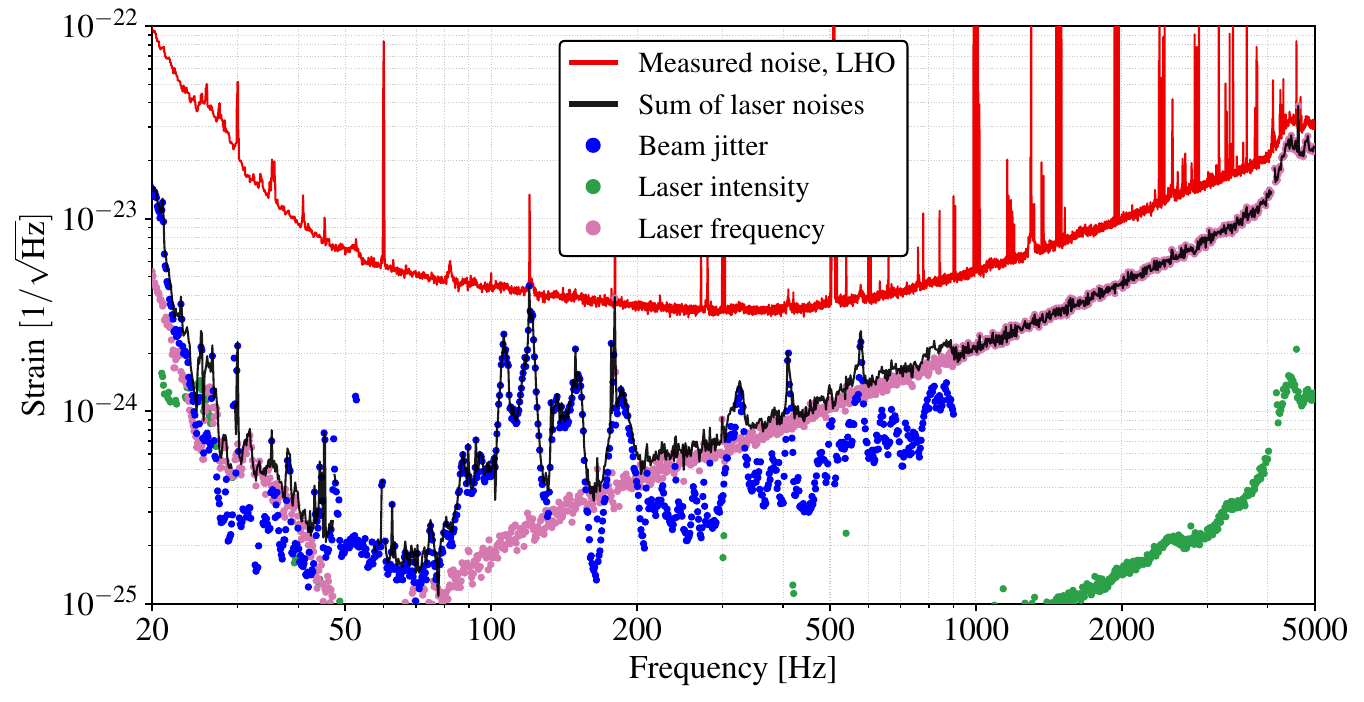}
    \includegraphics[width=0.48\textwidth, trim={0 0 0 0}]{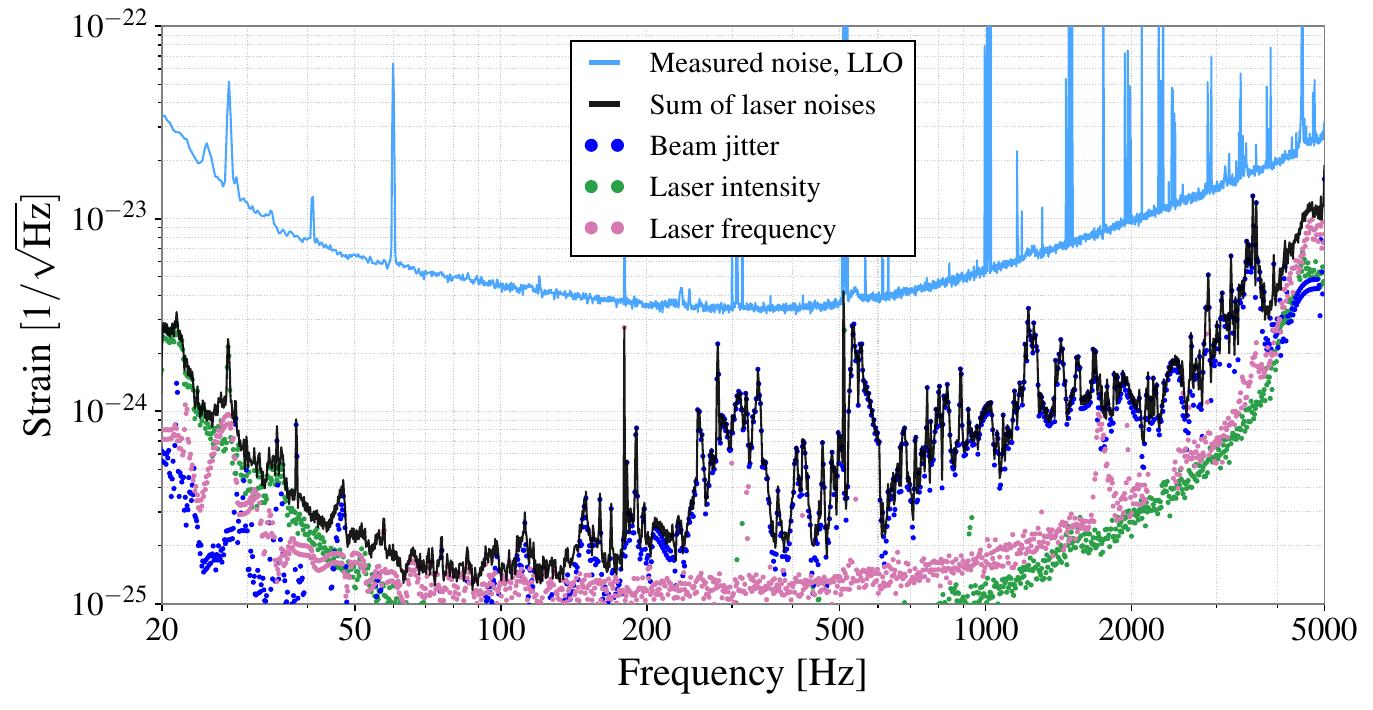}
    \caption{Laser noise sub-budget for the LIGO Hanford (upper) and Livingston (lower) detectors. Total laser noise includes the input beam jitter noise, laser intensity noise, and laser frequency noise. The input beam jitter noise and frequency noise dominate the laser noise, and the laser intensity noise is a limiting noise source below 40 Hz in the Livingston detector.
    }
    \label{fig:laser_subbudget}
\end{figure}

Frequency noise, intensity noise, and spatial fluctuations of the laser (beam jitter) also couple to the gravitational-wave readout~\cite{Izumi_FreqRespPart3_2015,Somiya:2006kb,Camp:2000ht,Cahillane:21}. Generally, frequency noise and intensity noise are described as sidebands on the main laser carrier frequency. Beam jitter can similarly be represented as sidebands of the Hermite-Gaussian first order mode~\cite{Morrison:94}. These noise sidebands enter the interferometer from its input port; ideally, in a perfectly symmetric interferometer, these sidebands are reflected back to the input port and do not appear at the anti-symmetric port where gravitational-wave signals are detected. This is known as common-mode rejection. In practice however, asymmetries exist due to e.g, the macroscopic arm length differences required for interferometer control, and differences in the mirror reflectivities, losses, power buildups, and transverse mode content. These asymmetries decrease the efficiency of common-mode rejection and allow the noise sidebands to couple with gravitational-wave signals. Therefore, it is necessary to stabilize the input laser frequency and intensity noise, and to mitigate the input beam jitter. \cref{fig:laser_subbudget} shows these contributions to the total laser noise for both detectors.

\subsubsection{Laser frequency noise} \label{sec:laserFreqNoise}

\begin{figure}[t]
    \centering
    \includegraphics[width=0.49\textwidth, trim={0 5mm 0 0cm}]{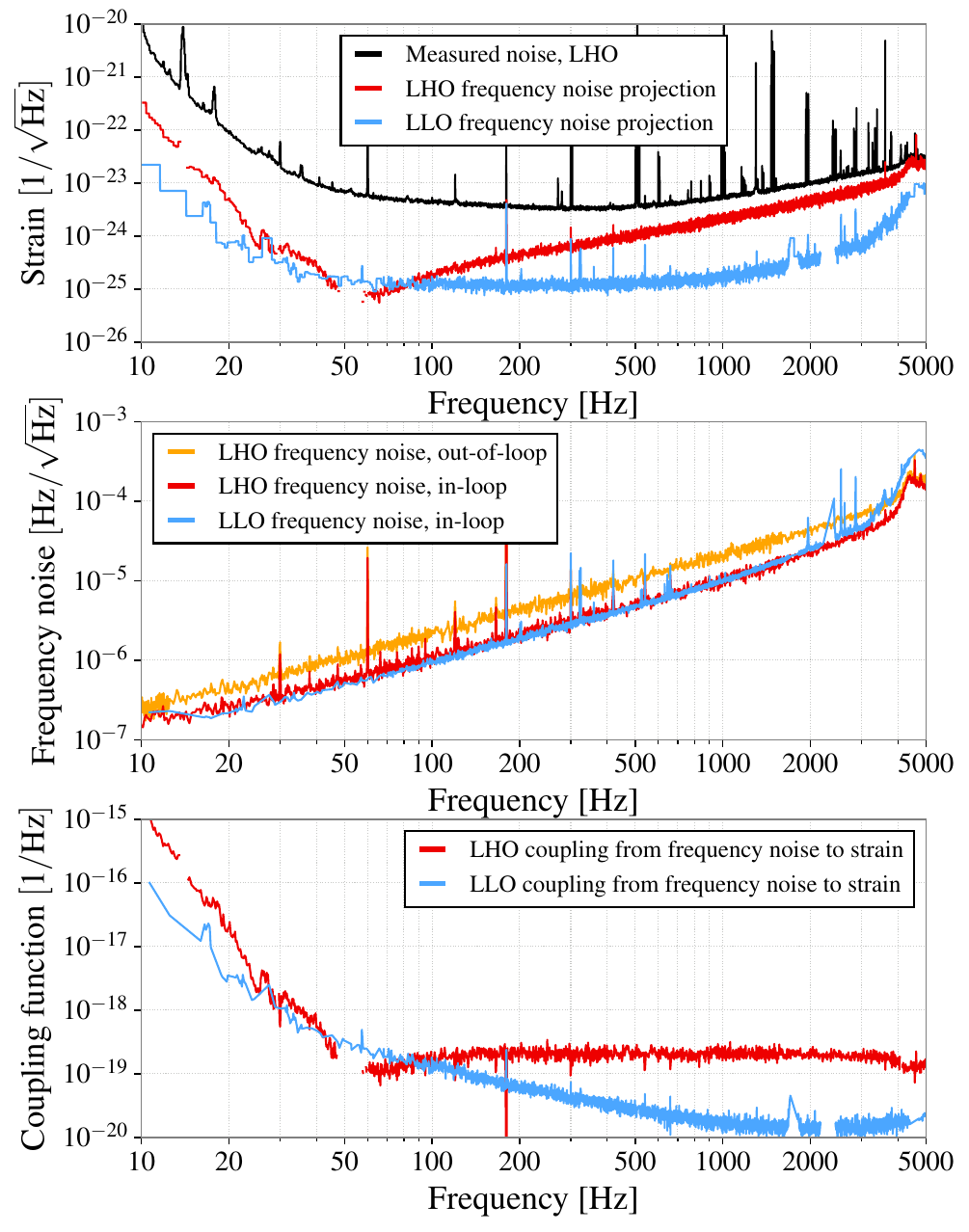}
    \caption{Laser frequency noise. The top plot shows the projected frequency noise for both detectors, compared to the total noise of the Hanford detector. The middle plot shows the calibrated frequency noise, as measured using the 4-km arm cavities as the reference. The Hanford detector has an out-of-loop sensor for the frequency stabilization loop to obtain both the in-loop and out-of-loop frequency noise. The Livingston detector has only the in-loop sensor. The bottom plot shows the coupling function from laser frequency noise to total strain noise.}
    \label{fig:freq_noise}
\end{figure}

The laser frequency is stabilized using three optical cavities as the frequency references. The first reference is a \qty{20}{cm} linear reference cavity made of ultra-low-expansion glass, located on the in-air optical table~\cite{Kwee2012, PSL2}. The second optical cavity employed as a frequency reference is the input mode cleaner, a \qty{33}{m} triangular cavity comprising three suspended mirrors. The third frequency reference is the \qty{4}{km} arm cavity length. While the motion of the differential arm length is utilized for gravitational-wave detection, the motion of the common arm length is employed as the ultimate frequency reference~\cite{Fritschel:01}. Errors from resonant frequencies of these references are fed back to the laser frequency through the nested feedback loop. The common arm length, input mode cleaner, and reference cavity stabilize the laser frequency in different frequency bands: below \qty{10}{\kHz}, between \qty{10}{\kHz} and \qty{50}{\kHz}, and \qty{50}{\kHz} to \qty{300}{\kHz}, respectively. 

\cref{fig:freq_noise} shows the frequency noise and its sensitivity projection at each site. The frequency noise is calibrated by the error signal of the common arm length stabilization loop, resulting in similar noise levels at both sites. However, the frequency noise coupling differs between the two sites due to higher order mode coupling caused by differences in the thermal states of the interferometers. When the curvature of the test mass is changed by the thermal compensation system, the higher-order transverse mode content in the arm cavities can change, which in turn can change the frequency noise coupling significantly~\cite{Cahillane:21}. This not only affects the coupling level but also alters the response shape, resulting in variation in the frequency and width of the dip around 50~Hz in the Hanford detector's coupling function in \cref{fig:freq_noise}. Adjustments are made to the thermal compensation system to ensure that the frequency noise does not limit the sensitivity (see \cref{sec:tcs}), and that the detectors are thus maximally shot-noise-limited at kilohertz frequencies.

\subsubsection{Laser intensity noise} \label{sec:laserIntensityNoise}
Laser intensity noise is stabilized by two feedback loops~\cite{ISSPatrick}. The first loop involves a photodetector installed on an in-air pre-stabilized laser table, which is fed back to an acousto-optic modulator also installed on the same table. This loop operates within a bandwidth of \qty{80}{\kHz}. The second loop stabilizes the output power of the input mode cleaner with the same actuator as the first loop. Here, eight photodiode arrays installed in vacuum are used to stabilize the power with larger power than the first loop, effectively reducing shot noise, which limits the noise of the photodiode arrays. This loop has a bandwidth of \qty{30}{\kHz}. These loops stabilize the relative intensity noise to a level of $4\times10^{-9}/\sqrt{\text{Hz}}$ between \SIrange{30}{1000}{Hz}.

\subsubsection{Beam jitter noise} \label{sec:beamjitter}
Input beam jitter noise has been one of the most significant noise sources masking the gravitational-wave signal in O4. Currently, LIGO does not have an active stabilization system for beam jitter noise similar to other laser noises, but a passive reduction of jitter is achieved using the input mode cleaner. As mentioned above, beam jitter can be described as a transfer of energy from the fundamental Hermite-Gaussian mode to first-order modes~\cite{Morrison:94}, and in a cavity resonating at the fundamental mode, higher-order modes are filtered out, leading to a reduction in the input beam jitter in the output beam of the cavity~\cite{Mueller2005}.

However, the input beam jitter noise is significant, limiting sensitivity in multiple bands between \qtyrange{100}{1000}{\Hz} at Hanford and around \qty{500}{\Hz} at Livingston. Jitter noise coupling, like frequency noise, is influenced by the thermal state of the interferometer. At Hanford, the coupling of input jitter noise increased by a factor of about ten as the input laser power was raised from 46 W to 72 W, and decreased by a factor of about two when the power was lowered to 57 W. One hypothesis is that the higher power produces greater thermal distortions of test mass surfaces around coating defects, making the arms less balanced and reducing common-mode rejection of input noise~\cite{brooksPointAbsorbersAdvanced2021a}.  

\begin{figure}[t!]
    \centering
    \includegraphics[width=0.49\textwidth, trim={0 0mm 0 0cm}]{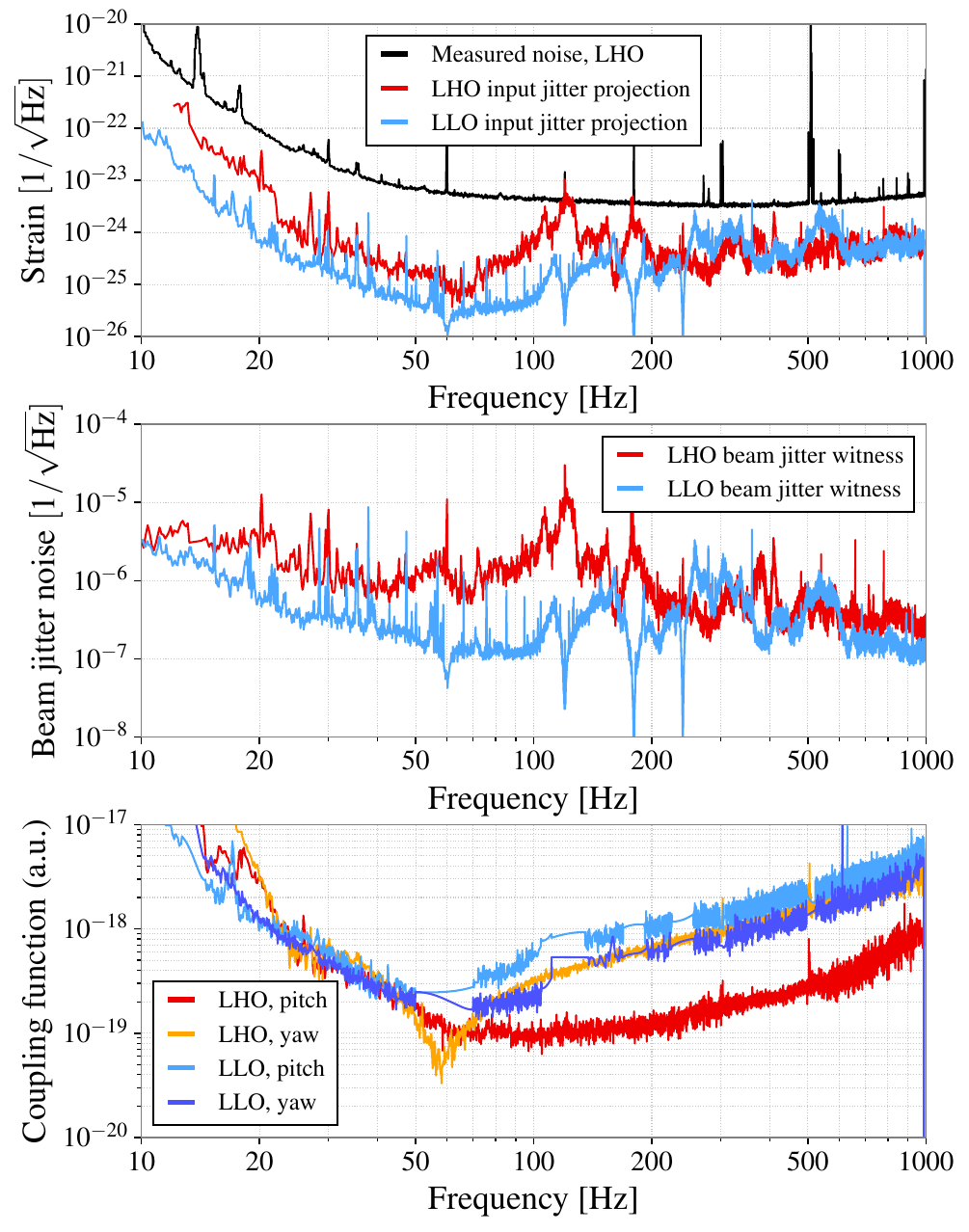}
    \caption{Strain projection of the noise from input beam jitter (top), quadrature sum of the input beam jitter noise of pitch and yaw direction  (middle), and the coupling function from input beam jitter noise to gravitational-wave strain noise (bottom). The jitter noise is shown with the unit of relative amplitude fluctuation of the 1st order Hermite-Gaussian mode to the fundamental mode. The sensor for the jitter noise is located at the reflection port of the input mode cleaner.
    }
    \label{fig:jittercoupling}
\end{figure}

Above \qty{100}{\Hz}, it can be seen from the bottom plot of \cref{fig:jittercoupling} that the coupling levels at Hanford and Livingston are similar, except for the pitch coupling at Hanford, which is not understood. However, the gravitational-wave strain projection of the input beam jitter noise at Hanford is about an order of magnitude larger than at Livingston around \qtyrange{100}{300}{\Hz}. This is because of the higher environmental noise in the laser room at Hanford compared to Livingston, as seen in the middle panel of \cref{fig:jittercoupling}, which witnesses the jitter on wavefront sensors in reflection of the input mode cleaner (\cref{fig:ligo}). This jitter is largely introduced by vibrations and resonances of the periscope that relays the main input beam from the optical table in the laser room to the interferometer plane. For O4, using these witness sensors in input mode cleaner reflection, much of the linearly coupled component of this jitter can be subtracted online in low-latency using the online noise cleaning described in \cref{sec:onlinecleaning}.

\subsection{Controls noise} \label{sec:controlnoise}

In addition to the control of the \FP{} cavity differential arm length from which the gravitational-wave readout is derived, the lengths and alignment of the arms and auxiliary cavities must be controlled to ensure stable resonance of light in each cavity. The majority of this control is performed through signals derived from the electro-optic modulation of the main interferometer beam, generating three radio frequency sidebands (see \cref{fig:ligo}) that are used for sensing each degree of freedom. Other methods of length and alignment control used in LIGO include dither schemes, DC pointing schemes, and camera servos. In addition to this ``global control'' that relies on controlling detector degrees of freedom, ``local control'' on each mirror suspension also suppresses excess mirror motion, sensed and controlled by local Optical Sensors/ElectroMagnetic actuator units (OSEMs) mounted on the suspension~\cite{Aston2012,BOSEMs}.

The gravitational-wave readout is marginally sensitive to other length and alignment degrees of freedom beyond the differential arm length, and the control system design is optimized around maintaining a stable lock point for all degrees of freedom while suppressing noise contamination. This system is not perfect, and \cref{fig:noisebudgets} contains several traces that demonstrate the coupling of various types of controls noise to the gravitational-wave strain, generally dominating at low frequency. In O3, both detectors were directly limited by length and alignment controls noise below 30 Hz~\cite{o3paper}; between O3 and O4, there was a significant effort to further optimize the length, alignment, and local suspension controls to increase low frequency sensitivity. 
Improvements in sensors, redesign of control loops, and novel control methods commissioned for the fourth observing run led to a factor of ten reduction in alignment controls noise and a factor of two reduction in auxiliary length controls noise near 20 Hz at both detectors.

\subsubsection{Auxiliary length control noise}\label{sec:lengthcontrolnoise}

Auxiliary lengths include the Michelson cavity length, formed between the beamsplitter and the two input test masses; power-recycling cavity length, formed between the power-recycling mirror (PRM) and the average distance of the input test masses; and signal-recycling cavity\footnote{While this coupled cavity is historically referred to as the ``signal-recycling cavity'', it is actually configured for \emph{resonant sideband extraction} that broadens the detection bandwidth~\cite{Mizuno:1993cj, bncPhysRevD.64.042006}. More recent papers appropriately call this cavity the ``signal-extraction cavity.'' Following historical practices, this cavity and its mirror components will be referred to as ``signal-recycling'' in this paper.} length, formed between the signal-recycling mirror (SRM) and the average distance of the two input test masses (refer to \cref{fig:ligo} for a detector schematic). These lengths are controlled from signals derived at the pick-off port (POP) of the power-recycling cavity, indicated in \cref{fig:ligo}.
The \cref{fig:noisebudgets} ``auxiliary length control'' traces demonstrate the measured noise contribution from these three auxiliary lengths at both detectors. Despite careful length control design to minimize noise re-injection within the gravitational-wave band, these auxiliary length controls still imprint excess sensing noise. To reduce the noise contribution of these loops to the gravitational-wave signal, a feedforward cancellation is applied.

\begin{figure}[t!]
    \centering
    \includegraphics[width=0.5\textwidth,trim={0 0 0 0}]{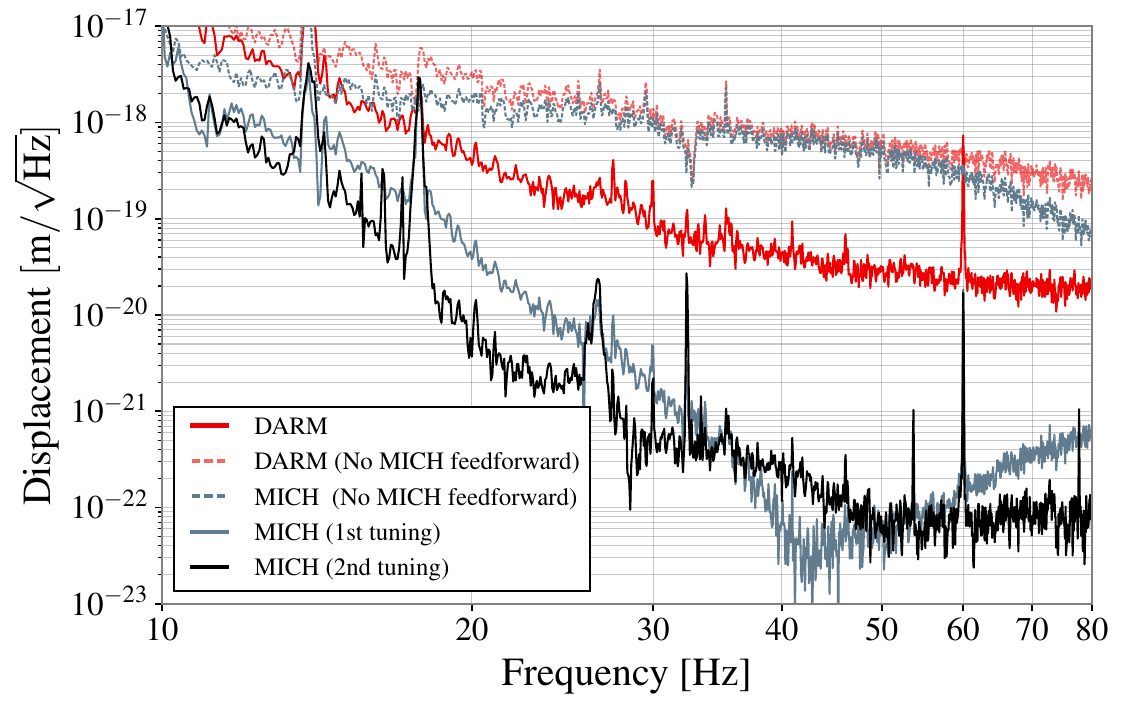} 
    \caption{A comparison of the differential arm length (DARM) and Michelson length (MICH) coupling at LIGO Hanford. With no feedforward applied, coupled Michelson noise dominates the differential arm length at low frequency. With an initial feedforward tuning, this coupled noise is suppressed to no longer limit total detector noise. Iterative tuning of the feedforward provides a better approximation of the noise coupling, allowing the feedforward to suppress it further.
    }
    \label{fig:lsc_subbudget}
\end{figure}

Feedforward cancellation is a method of real-time noise cancellation by mechanically driving the expected sensing noise to the test masses. It is made possible by the fact that the residual noise of the auxiliary lengths is witnessed by independent sensors from the gravitational-wave readout. The transfer functions from each auxiliary loop to differential arm length are each measured, as well as the transfer function of the feedforward actuation path. The transfer functions are fit using algorithms such as \texttt{vectfit} or IIRrational \cite{vectfit, iirrational}. The opposite sign of the coupling of this auxiliary loop is injected into the differential arm length using a fit of the measured transfer function to cancel out the noise that would appear. This feedforward scheme can reduce the contribution of auxiliary length noise in the differential arm length by a factor of 100 or more in the gravitational-wave band. \cref{fig:lsc_subbudget} shows how the noise contribution of the Michelson length motion changes with feedforward control applied.

There are a few limitations to this method related to the measurement quality and ability to fit the measured transfer function.
To overcome this challenge, an iterative method of feedforward tuning was applied at the Hanford observatory~\cite{LSCFFiterative}. After initial injection, an appropriate feedforward filter was tuned to subtract the excess noise. Next, the same injection was applied, with the new feedforward engaged. This allowed any residual linear coupling to be measured and suppressed further with an improved filter, with each iterative tuning step providing a better approximation of the true transfer function. \cref{fig:lsc_subbudget} compares Michelson noise coupling to differential arm length in each of these scenarios measured at LIGO Hanford. A comparison between the first and second tuning trace shows that overall, the iterative method improves measurement precision enough to allow further noise subtraction. However, around 40 Hz, some worsening of the coupling is witnessed due to transfer function fitting limitations. This iterative method was applied for both the Michelson and signal-recycling cavity feedforward tuning at Hanford.

Feedforward cancellation was applied to suppress the noise from the Michelson and signal-recycling cavity loops in O4. While noise from the power-recycling cavity also couples, this coupling is usually indirect, and results from improper sensor diagonalization or other opto-mechanical coupling to other degrees of freedom. Therefore, it can be reduced through the main feedforward application to the Michelson and signal-recycling cavity lengths. With the iterative tuning method, a factor of one hundred or more suppression of Michelson length noise is possible, as is a factor of thirty suppression of signal-recycling cavity length noise.

\subsubsection{Alignment control noise} \label{sec:alignmentcontrolnoise}

\begin{figure}[t]
    \centering
    \includegraphics[width=\columnwidth]{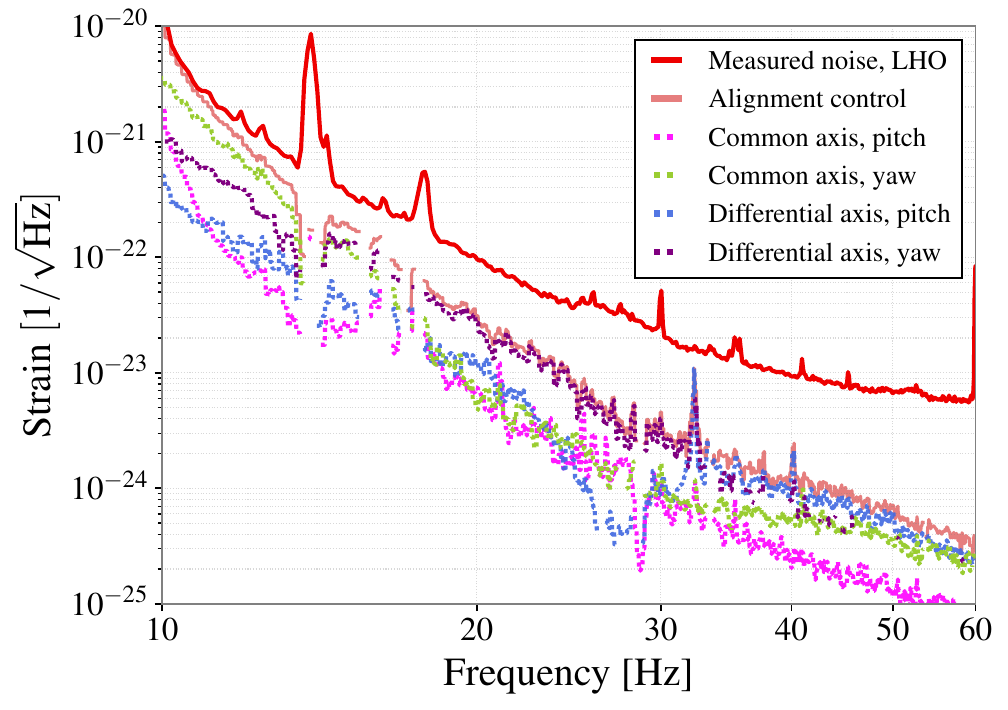}
    \caption{A sub-budget showing the most-limiting alignment control loops measured at the Hanford observatory, the four arm cavity ``HARD'' loops~\cite{BarsottiASC}. These four loops control the arm cavity axis alignments using sensors at the REFL and AS ports (see \cref{fig:ligo}), and have bandwidths between 3 and 5 Hz.}
    \label{fig:asc_subbudget}
\end{figure}

The most significant alignment coupling results from the control of the arm cavity alignment, as any change in arm cavity mirror alignment is a direct change in arm length. Eight control loops govern the common and differential cavity axis alignment and cavity pointing in pitch and yaw; referred to as ``HARD'' loops for cavity axis alignment, and ``SOFT'' loops for cavity pointing~\cite{BarsottiASC}. The cavity axis alignment is controlled with bandwidths of up to \qty{5}{Hz}, and the cavity pointing alignment with lower bandwidths that are no more than \qty{1}{Hz}. \cref{fig:asc_subbudget} shows the four most limiting alignment control loops, the HARD loops, and their contributions to gravitational-wave strain at LIGO Hanford. These loops are controlled from wavefront sensors at the symmetric (REFL) and antisymmetric (AS) ports of the interferometer, indicated in \cref{fig:ligo}~\cite{BarsottiASC}. Global alignment control is also implemented to govern the alignment of the auxiliary cavities, using bandwidths that range from a few milliHertz to a few Hz. These control loops also couple noise to the gravitational-wave readout, but with coupling strengths that are an order of magnitude or more lower than the coupling of the arm cavity alignment.

While most wavefront sensors are sensing-noise-limited above 10\,Hz, at both Livingston and Hanford, wavefront sensors mounted on a passively damped in-vacuum table at the symmetric port are contaminated by excess noise from \qtyrange{10}{30}{\Hz} due to ground motion coupling. Feedforward subtraction is applied to reduce this noise coupling using accelerometers mounted on the table that witness the table motion. The feedforward is applied to subtract noise from all alignment control loops sensed at the symmetric port, including the control of the common hard mode of the arm cavities, power-recycling cavity alignment, and the input pointing. The feedforward can suppress the excess noise by up to a factor of ten.

Another significant source of alignment noise in O3 resulted from the alignment dither system used to control the beam spot positions on the mirrors in the arm cavities. This involved injecting a sine wave into the angular motions of the mirrors in the arm cavities, and then deriving a feedback control signal by demodulating the differential arm length signal at the dither frequencies, around \qty{20}{Hz} at Hanford and around \qty{10}{Hz} at Livingston~\cite{o3paper}.
To avoid the presence of dither lines in the gravitational-wave band in O4, the spot position control was moved to a camera servo system: video cameras monitor the test mass mirror surfaces, and the centroid of each image is calculated to estimate the beam spot position. At both observatories, the beam spot positions on the end test masses and the beamsplitter are monitored by cameras and fed back to control the angular motions of the test masses and the power-recycling mirror. The alignment dither system is still used for lock acquisition at the Hanford observatory, but the control is transitioned to camera servos before entering observing mode.

At the Hanford observatory, it is necessary to mis-center the beam spots on several arm cavity optics to avoid point defects on the test masses that can increase interferometer losses due to excess absorption~\cite{brooksPointAbsorbersAdvanced2021a}. The point defects are large enough that the beams are miscentered by several millimeters relative to the centers of the mirrors. The Livingston observatory no longer has significant point defects, so the beams are kept near the centers of the mirrors.

\subsubsection{Local suspension control noise}\label{sec:bosemnoise}

Each suspension is controlled locally to suppress excess motion, namely around the suspension resonances, using type A and B OSEMs~\cite{Aston2012, BOSEMs}.
The local control of each suspension is an AC-coupled control design at the top suspension stage only with frequency cutoffs below \qty{10}{Hz} to avoid sensor noise contamination in the gravitational-wave band. However, even with these cutoffs, OSEM sensor noise still contaminates the gravitational-wave strain.

\begin{figure}[t]
    \centering
    \includegraphics[width=\columnwidth]{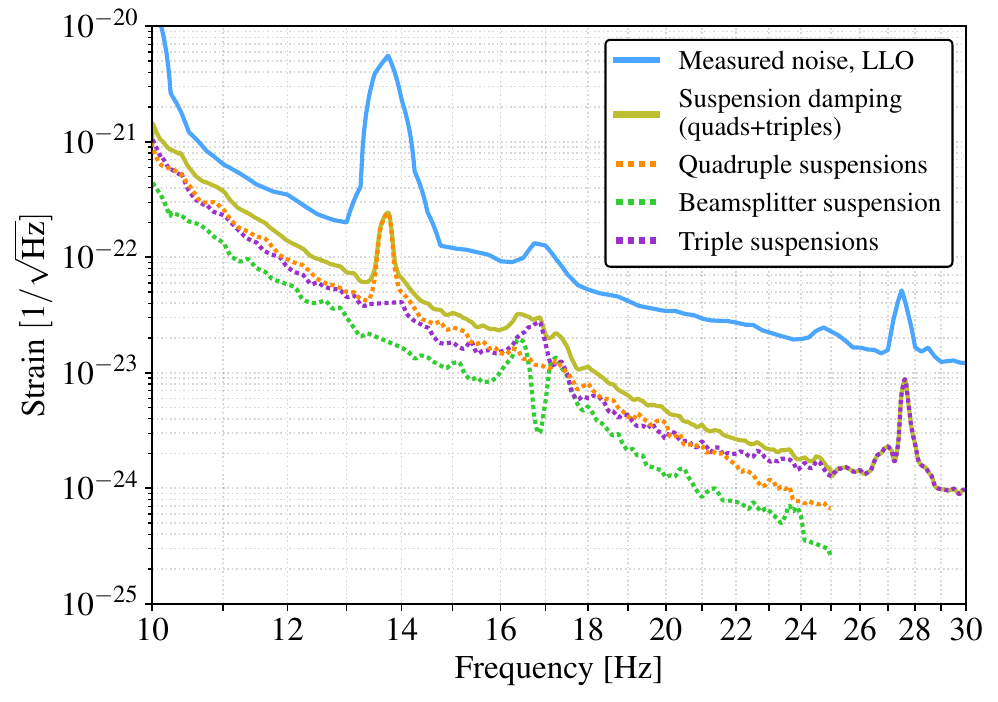}
    \caption{Local damping noise contributions to gravitational-wave strain noise measured at the Livingston observatory. After local control upgrades, contribution from auxiliary optic triple suspensions is comparable to the contribution from test mass quadruple suspensions.}
    \label{fig:damping_noise}
\end{figure}

The contribution of these local damping controls is shown in the \cref{fig:noisebudgets} noise budget traces labeled ``Suspension damping.'' The Livingston noise budget shows contributions from the quadruple suspensions (``quads'')  that control the test masses and the 
triple suspensions that control the recycling cavity mirrors and beamsplitter (``triples''). The Hanford budget shows contributions from only the quadruple suspensions. Both detectors experience noise contamination from the quadruple suspensions and all triple suspension types, but the direct contribution of the triple suspensions to the total gravitational-wave strain noise has only been measured at the Livingston detector. A sub-budget of the contributions from the quadruple suspensions and triple suspensions at the Livingston detector is shown in \cref{fig:damping_noise}.

At both detectors, the local controls designs were optimized to prioritize noise suppression above \qty{10}{Hz}, while maintaining sufficient suppression of suspension resonances. Some local controls are relaxed further when the full global alignment control is active, as interferometric control of the mirrors suppresses some degrees of freedom using lower sensing noise than the OSEMs.
At both sites, this effort was focused on quadruple and triple suspensions (test masses and recycling cavity mirrors). At the Hanford observatory, additional effort was applied to optimize the controls of single-stage suspensions along the input of the interferometer, between the input mode cleaner and the power-recycling cavity (see \cref{fig:ligo}).
At Hanford, these efforts corresponded to a reduction in local controls noise contamination of an order of magnitude or more above 10 Hz in the alignment controls, and a direct reduction of the gravitational strain noise by a factor of two between 20 and 30 Hz.

Despite these control optimization efforts, noise from the suspension controls still challenges the efforts to reduce low frequency noise. \cref{fig:damping_noise} shows that noise from all suspensions, not just the test mass suspensions, can directly limit detector sensitivity. Also, many of the length and alignment control loops, such as those depicted in \cref{fig:asc_subbudget}, are limited by noise from the OSEMs, and indirectly couple local controls noise into the gravitational-wave strain.

\subsection{Digitization noise}\label{sec:dacnoise} 

The digital-to-analog converters (DACs) used for the test mass control signals also contribute noise to the gravitational-wave strain. This digitization noise results from nonlinear behavior in the conversion of a discrete digital signal to a continuous analog signal~\cite{Widrow_Kollár_2008}. The most significant source of this noise in the LIGO detectors is from the penultimate-mass actuators on each test mass, included in \cref{fig:noisebudgets}.

Digitization noise can be reduced through a technique called dithering, which involves injecting high-frequency sinusoidal signals through the DACs~\cite{Widrow_Kollár_2008}. A large, high-frequency signal dominates the RMS of the DAC, such that the digitization noise at low frequency, where the key portion of the control signal is generated, is reduced.
At the Livingston detector, DAC noise limited the strain sensitivity at low frequency, therefore a single dither line at 7 kHz was injected and improved digitization noise by a factor of three, suppressing DAC noise sufficiently far below the total detector noise. 

While a similar digitization noise is observed at the Hanford detector, it is not a limiting noise, and the dithering method did not change the total displacement noise level.
Therefore, Hanford operates without DAC dithering.

\subsection{Photodetector dark noise} \label{sec:darknoise}

The photodetector dark noise trace in \cref{fig:noisebudgets} is the projected impact of total noise measured with no light incident on either of the two output photodiodes. 
This dark noise measurement includes noise from the entire electronics readout chain, such as the leakage dark current of the two output photodiodes, the Johnson--Nyquist noise of the readout transimpedence network, and all associated downstream readout electronics. 
In O4, dark noise lies approximately a factor of ten in amplitude below the total detector noise due to improvements along the entire signal chain. 

The first stage of the photodetector signal chain is their transimpedance amplifiers.
Transimpedance amplifier stages are limited by op-amp output voltage range; the maximum signal-to-noise ratio of the photodetectors used to sense the gravitational-wave signal is set by the photocurrent across the feedback resistor, previously dominated by the comparatively large DC component, $\sim 10$ mA. 
Between O3 and O4, a redesigned photodetector transimpedance network was installed based on the concept demonstrated at GEO600~\cite{GroteLowNoisePhotodetector}. 
The network includes a preceding current divider; a \qty{2.5}{\henry} inductor connected in parallel with two \qty{10}{uF} capacitors and \qty{3}{k\Omega} resistor to ground, which creates a passive, second-order, RLC, high-pass filter with a 25 Hz corner frequency. 
This diverts the majority of photocurrent below 10 Hz to ground, leaving the audio-band transimpedance amplifier only 1\% of the photocurrent -- but that which contains all the interesting gravitational-wave signal. 
This topology~\cite{DCPD_readout} allows for improved signal-to-noise ratio and higher dynamic range, an improvement of a factor of five over O3 ~\cite{o3paper}. 

Additional upgrades further down the photodetectors' readout chain were also made to ensure that the transimpedance stage remains the dominant source of electronics dark noise in the gravitational-wave frequency band in O4. 
The analog-to-digital converter (ADC) cards were updated to General Standards 66-18AI64SSC750K cards,
reducing noise in each ADC channel by a factor of five. 
As such, no analog whitening is needed in the interferometer's lowest noise configuration, making the noise introduced by whitening negligible.
The upgraded ADC cards also changed the native sampling rate from 65 kHz to 524 kHz, requiring a rearrangement of timing and data exchange in the real-time front-end software system, as well as updated analog and digital anti-aliasing filters to support the faster sample rate. 
At the Hanford observatory only, each of the two photodetectors' analog signal voltage was copied and mapped to four ADC channels and averaged digitally, reducing coherent noise on each channel by an additional factor of two.

\subsection{Online noise subtraction} \label{sec:onlinecleaning}

In O4, up to three calibrated strain channels from each detector were available: calibrated strain, calibrated strain with low-latency line subtraction, and calibrated strain with low-latency line subtraction and noise subtraction\footnote{These channels are named \texttt{GDS-CALIB\_STRAIN}, \texttt{GDS-CALIB\_STRAIN\_NOLINES}, and \texttt{GDS-CALIB\_STRAIN\_CLEAN}, respectively.}. \cref{fig:rangeplots} reports the range with both online line subtraction and noise cleaning for LIGO Hanford, and with online line subtraction for LIGO Livingston. \cref{fig:noisebudgets} show the strain sensitivity for both detectors with line and noise subtraction applied.

Both detectors provide the calibrated strain with low-latency line subtraction, where the sinusoidal calibration lines and the 60 Hz power lines are removed from the online data~\cite{Viets:2021aaa}. The line subtraction provided over 2 Mpc of additional BNS range for both detectors.

The Hanford observatory, in addition to low-latency line subtraction, implemented online ``cleaning'' of the strain data to subtract residual noise couplings during O4~\cite{Vajente_nonsens, Mukundbilinear}. This data cleaning is performed using the NonStationary Estimation and Noise Subtraction code (NonSENS)~\cite{Vajente_nonsens,NonSENS_tech_report,nonsens_code}. NonSENS has previously been demonstrated to successfully subtract both linear stationary couplings and nonlinear modulated couplings in the offline data from O3~\cite{PhysRevD.105.102005}. For O4, front-end infrastructure was developed to perform cleaning of the strain noise data in low-latency using measured noise couplings.

The greatest benefit of this online cleaning is in subtracting laser noises such as beam jitter, described in \cref{sec:beamjitter}, and frequency noise, described in \cref{sec:laserFreqNoise}. The noise reduction with online cleaning could increase the BNS range by up to 3 Mpc for Hanford due to its mitigation of nonstationary laser noise couplings. 
Together, for Hanford, the online line subtraction and noise cleaning improved the cleaned BNS range by up to 5 Mpc.
The Livingston observatory also implemented online noise cleaning for O4 to subtract jitter noise, resulting in a 1-2 Mpc improvement in the cleaned BNS range. 

\subsection{Correlated noise}\label{sec:corrnoise}

The gravitational-wave readout is performed by two photodiodes that detect the two equal-power beams in the transmission of the output mode cleaner (\cref{fig:ligo}). As a result, the strain noise spectra for the gravitational-wave readout is given by the amplitude spectral density of the sum of the two output photocurrents; this summed photocurrent noise spectra is calibrated into strain noise, and shown in the detector noise budgets of \cref{fig:noisebudgets}. 
To more deeply study the detector noise, the correlated noise between the two output photocurrents can be measured. Displacement noise terms are correlated between the two photodetectors, while sensing noises such as photodetector dark noise and shot noise are uncorrelated and thus average out of the cross power spectrum when quantum squeezing is not injected~\cite{IzumiXcorr2017}.
Therefore, measuring the correlated noise between the two output photodetectors without squeezing can reveal the underlying noises below shot noise, such as coating thermal noise \cite{PhysRevA.95.043831}.

\begin{figure*}[] 
    \centering
    \begin{minipage}{0.49\linewidth}
        \includegraphics[width=\linewidth]{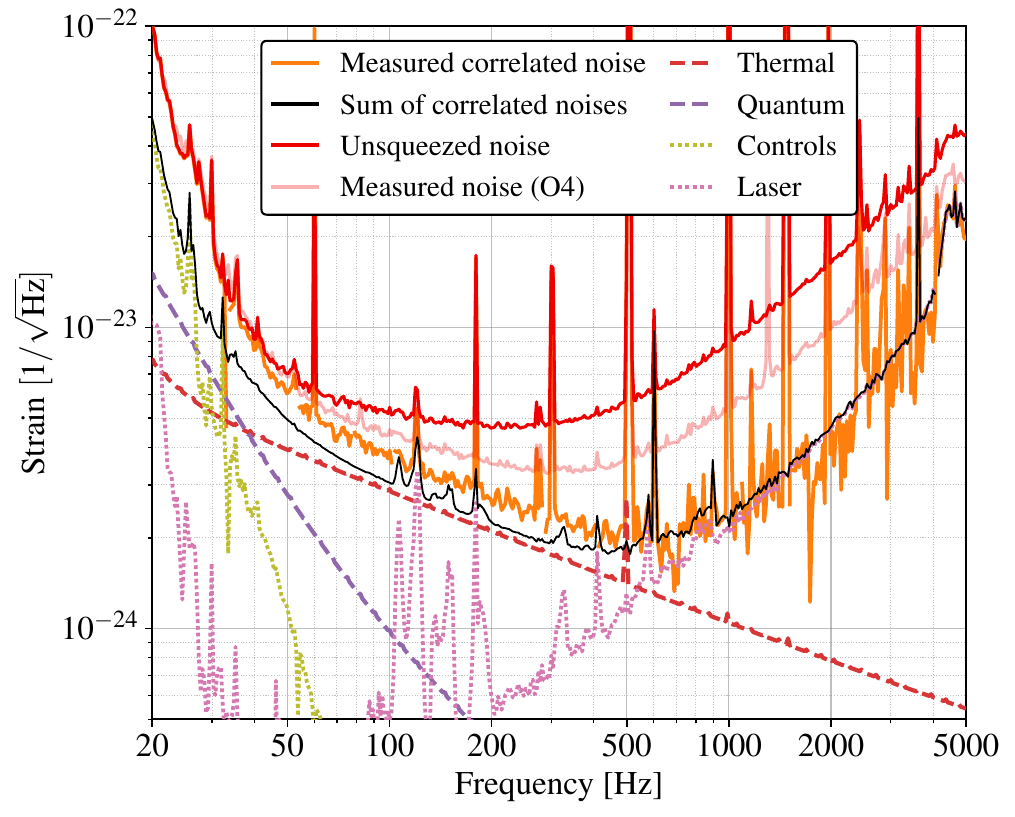}
        \par{(a) LIGO Hanford Observatory.
        } 
    \end{minipage}
    \begin{minipage}{0.49\linewidth}
        \includegraphics[width=\linewidth]{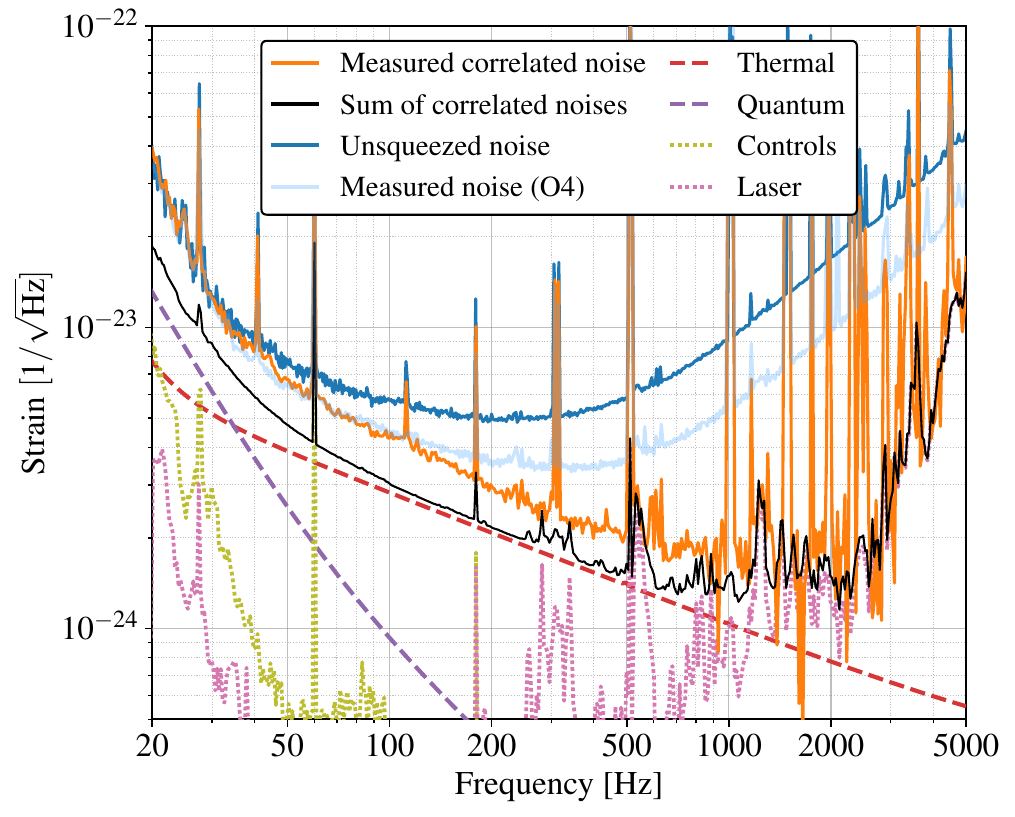}
        \par{(b) LIGO Livingston Observatory. 
        } 
    \end{minipage}
    \caption{Correlated noise budgets for the LIGO Hanford (left) and Livingston (right) detectors. 
    Orange traces show the correlated strain noise amplitude spectral densities measured from the two readout photodiodes~\cite{IzumiXcorr2017}, for a time without squeezing injected.  
    The total unsqueezed detector noise spectra for these times is shown in red (Hanford) and blue (Livingston) traces. The light traces (``Measured noise (O4)'') show the total detector noise spectra with squeezing injected, as in \cref{fig:noisebudgets}. Black lines show the sum of expected correlated noises. Dashed lines show the calculated total thermal noise (red, Sec.~\ref{sec:ctn}) and the estimated quantum radiation pressure noise (purple) given the detector readout angle and detunings (Sec.~\ref{sec:sqz_characterization}). Dotted lines show the total expected controls noise (olive, which is the sum of alignment, length, suspension damping, and actuator controls noises described in Sec.~\ref{sec:controlnoise}), and laser noise (pink, which includes laser intensity, frequency, and jitter noises as in Sec.~\ref{fig:laser_subbudget} and Sec.~\ref{sec:lasernoise}). 
    The correlated noise spectra use the mean-averaged unsqueezed detector noise power spectral densities of the two output photodiodes averaged over 900 seconds for Hanford and 1800 seconds for Livingston.
    }
    \label{fig:corr_noise_budget}
\end{figure*}

\cref{fig:corr_noise_budget} shows the correlated noise budgets for both detectors, for a time without squeezing injected. 
Above \SI{1}{\kHz}, secondary to shot noise, the correlated noise budgets indicate that the laser noises described in \cref{sec:lasernoise} limit detector sensitivity. At Hanford, the correlated noise appears well-explained by the laser frequency noise projection. At Livingston, the correlated noise appears well-explained by the input beam jitter noise projection.

In the mid-band between 100 Hz and 500 Hz, coating thermal noise is expected to be the primary noise source (see \cref{sec:ctn}). 
The correlated noise budgets shown in \cref{fig:corr_noise_budget} are estimated using the mean-averaged power spectral densities, which will include contributions from temporally stationary noises such as coating thermal noise (\cref{sec:ctn}), as well as non-stationary noises such as input beam jitter (\cref{sec:beamjitter}) or scattered light noise (\cref{sec:scatteredlight}). 
Measurements using mean-averaged spectra indicate a discrepancy between the measured noise and the expected noise, which is visible in both the total squeezed noise budgets (\cref{fig:noisebudgets}) and the unsqueezed correlated noise budgets (\cref{fig:corr_noise_budget}). This noise floor appears nearly unaffected by experimental changes in the interferometer power or injected squeezing. Additionally, the discrepancy appears larger for the Livingston detector than the Hanford detector. Moreover, at Livingston, the amplitude of this excess noise was observed to decrease with time after the installation of new end test masses during the mid-run break between O4a and O4b (\cref{sec:testmasses}).

\section{Instrument upgrades}\label{sec:improvements}

Major detector upgrades were implemented between O3 and O4 to target noise reduction and improve detector operation.
Upgrades to the pre-stabilized laser system increased available input power. Test mass replacements rectified problems with nonuniform coatings and excess loss. 
Scattered light noise was reduced through targeted removal or adjustment of known scattering surfaces. Electromagnetic interference was reduced through adjustment of electronics grounding.
As a first stage of the A+ upgrade, the squeezer system was upgraded to inject frequency-dependent squeezing, and system-wide losses were reduced to enable higher squeezing levels. This upgrade covered an extensive array of improved or new squeezer and output optics components, including new active mode-matching mirrors, faraday isolators, and new vacuum and seismic isolation systems to support a 300 m filter cavity.

Between O4a and O4b, a 3-month commissioning break from January 16, 2024 to April 10, 2024 allowed the observatories to install several upgrades to improve sensitivity and operation. 
At Livingston, both end test mass mirrors were cleaned with First Contact \emph{in-situ}. At Hanford, the output mode cleaner cavity was replaced to reduce losses, baffles near the input mode cleaner were adjusted and damped to reduce scattered light noise, and the squeezer's piezo-deformable optics were adjusted to allow further optimization of squeezer mode-matching. At both detectors, a pre-mode cleaner cavity was also installed for the squeezer laser, with the goal of reducing RF laser intensity noise on the squeezer's control sidebands to improve squeezer alignment controls. Further controls optimization at the Hanford observatory reduced low frequency nonstationary noise from the differential arm length actuators.

\subsection{Pre-Stabilized Laser upgrade}\label{sec:psl}

The Pre-Stabilized Laser (PSL) was upgraded at both observatories to supply higher input power for the interferometers. The PSL begins with a 2 W NPRO 1064 nm seed laser. The seed beam is passed through 2 neoVAN-4S-HP amplifiers, the first amplifying the beam to 70 W and the second amplifying to 140 W of power. The amplified laser beam passes through an optical cavity called the pre-mode cleaner for spatial mode cleaning and filtering of intensity noise in the radio frequency (RF) band. The laser frequency is actively stabilized by using a rigid optical cavity as the frequency reference. The intensity is also actively stabilized using the photodetector and an acousto-optic modulator. The phase modulation for the main interferometer control scheme is applied to the PSL through an electro-optic modulator, as shown in \cref{fig:ligo}. 
The PSL beam path includes a rotation stage waveplate to control the input laser power injected into vacuum. The maximum power that can be supplied to vacuum is \qty{110}{\watt}. In O4, both observatories used between \qtyrange{60}{75}{\watt} of PSL input power during observing.

\subsection{Test mass replacements}\label{sec:testmasses}

Test mass mirrors were replaced at both observatories due to point defects in the mirror coatings, which contributed to control instabilities and excess noise from scattered light and beam jitter~\cite{brooksPointAbsorbersAdvanced2021a, o3paper, O3b}. These instabilities and excess noise also limited the maximum operating power of the detectors in previous observing runs~\cite{o3paper,O3b}.

At the Hanford observatory, the input test mass mirror to the Y-arm (ITMY) was replaced for O4. The presence of a point defect on ITMY was noted during the commissioning for O3, and nonuniform absorption from this point defect compromised the operation of the detector, especially when increasing the operating power from 25 W to 40 W \cite{o3paper}. 
After replacing ITMY, the Hanford observatory achieved a power-recycling gain (PRG) of 50\,W/W with an input power of 57\,W in O4, compared to a PRG of 44\,W/W with an input power of 34\,W input power in O3. 
No point absorbers have been observed on the ITMY mirror since its installation, according to Hartmann Wavefront Sensor images~\cite{Brooks:07,Brooks:16} of the mirror surface during high-power operation. However, point absorbers remain on the input test mass mirror to the X-arm (ITMX) and both end test masses.

At the Livingston observatory, both end test masses (ETMX and ETMY) were replaced ahead of O4, in order to remove point absorbers that limited the operation of the detector \cite{brooksPointAbsorbersAdvanced2021a}. 
After replacement, no point absorbers have been observed on either end mirror according to Hartmann Wavefront Sensor images of the ETMs. 
However, after installation of these new ETMs, the PRG in the Livingston detector decreased from 47 W/W with an input power of 38\,W in O3, to 35\,W/W with an input power of 63\,W in O4. This decrease in power-recycling gain is due to the increased cavity loss from the higher scatter seen from the new ETM coatings, and higher optical absorption in the mirrors.
Additionally, since the install of the new mirrors, the Livingston detector observed a higher total noise amplitude around 100 Hz in O4a, compared to in O3. This excess noise appeared unchanged for different squeezing levels and intracavity powers. 
The strain amplitude of the excess noise has decreased by approximately 20\% since the install of the mirrors before O4a. 
While this excess noise appeared in the frequency band dominated by thermal noise, and with a similar frequency dependence to that of coating thermal noise, the ETM witness samples coated alongside the ETMs exhibited the expected level of coating thermal noise (see \cref{eq:CTN})~\cite{amatoOpticalMechanicalProperties2021}, and not the elevated noise level observed in the full detector.

During the break between O4a and O4b, both Livingston ETMs were cleaned using First Contact. After this cleaning, wide-angle scatter from each mirror decreased around \qtyrange{20}{40}{\%}. Correspondingly, the power-recycling gain at Livingston increased by about 15\%.

\subsection{Squeezing upgrades} \label{sec:sqzUpgrades}
A major detector upgrade between O3 and O4 was the installation of a new \SI{300}{m} filter cavity for the injection of frequency-dependent squeezed vacuum; this upgrade has been detailed extensively in Refs.~\cite{ganapathyBroadbandQuantumEnhancement2023} and~\cite{jiasubSQL}. 
Other components in this upgrade have increased the generated squeezing levels, reduced losses, and improved mode-matching. Altogether, compared to the squeezing levels of 2-3 dB in O3~\cite{tsePRL19QuantumEnhancedAdvanced,mccullerPRD21LIGOQuantum}, squeezing levels of up to \SI{\sqzmaxlho}{dB} 
at Hanford and \qty{\sqzmaxllo}{dB} at Livingston have been observed in O4.

\emph{Filter cavity}. 
The filter cavity prepares frequency-dependent squeezed states of vacuum that go on to reduce the limiting form of detector quantum noise in each frequency band~\cite{ganapathyBroadbandQuantumEnhancement2023}: injecting amplitude-squeezed vacuum below $\sim$100 Hz to reduce quantum radiation pressure noise, and injecting phase-squeezed vacuum at kilohertz frequencies to reduce shot noise. This technique thus achieves broadband quantum noise reduction, allowing higher generated squeezing levels to be utilized.

\emph{More generated squeezing}. 
Both detectors used \qtyrange{11}{17}{dB} of generated squeezing (inferred) in O4, compared to \qtyrange{7}{9}{dB} in O3~\cite{tsePRL19QuantumEnhancedAdvanced,yuN20QuantumCorrelations,mccullerPRD21LIGOQuantum}, which contributed to the higher squeezing levels observed. For O4, more second harmonic green pump power was made available in-vacuum, thanks to updated in-vacuum green fibers with a higher damage threshold. In addition, two squeezer cavity mirrors were changed to lower the squeezer cavity threshold power, and thus reduce the green pump power required for operation.

\emph{Reduced losses}. Detector-wide losses were reduced to increase squeezing efficiency. Three new high-efficiency Faraday isolators with single-pass losses of \qtyrange{0.5}{1}\%~\cite{LIGOT1900788v2OutputFaraday} were implemented.  
Additionally, the output mode cleaner (OMC) cavity was replaced at both detectors. At the Livingston detector, the OMC was replaced before starting O4; the new OMC was estimated to have a throughput of $\sim$98\%.
At the Hanford detector, the OMC was replaced during the commissioning break between O4a and O4b; the new OMC is expected to have a throughput of $\sim$96\%.

\emph{Improved mode-matching}.
The measured squeezing is sensitive to the mode-matching, or the spatial overlap, between the squeezed laser beam, the fundamental spatial mode of the interferometer, and the output mode cleaner. 
To optimize mode-matching across the detectors, four new deformable mirrors with a controllable radius of curvature, known as Suspended Active Matching Stages (SAMS), were installed. 
On the squeezed light injection path, three SAMS controlled by piezoelectric actuators (``PSAMS''~\cite{srivastavaOEO22PiezodeformableMirrors}) were installed: one to mode-match the squeezed beam to the filter cavity, and two to mode-match the squeezed beam to the interferometer and output mode cleaner. 
On the interferometer readout path, a thermally-actuated SAMS (``TSAMS'' \cite{caoOEO20EnhancingDynamic,caoAOA20HighDynamic}) was installed to mode-match the interferometer beam through the output mode cleaner cavity. 
At the Hanford detector, during the commissioning break between O4a and O4b, the tunable range of two PSAMS optics in the squeezed light injection path was adjusted to improve mode-matching of the squeezed beam through the detector.

\subsection{Scattered light mitigation} \label{sec:scatteredlight}

As laser light propagates through the detectors, each optic surface can reflect a small fraction of the incident light and produce scattered light or stray beams, both of which can reflect off surrounding surfaces and re-interfere with the main interferometer beam. Motion of the scattering surfaces modulates the phase of the back-reflected scattered light, introducing parasitic phase modulations on the main laser beam~\cite{Vinetscatter, Ottaway:12}. 

Scattered light noise depends on the velocity of the scattering surface, and the amplitude ratio of the scattered light to the main laser light~\cite{Soni_2021}. The peak frequency at which scattered light noise is observed is proportional to the velocity of the vibrating surface, meaning that even if the amplitude and frequency of the vibrating surface is low, scatter can still produce noise within the observation band of gravitational waves depending on the vibration amplitude and the optical power in the stray beams. 
Due to the typical velocity of vibrating surfaces, scattered light impacts detector noise most at lower frequencies. 
From the noise budgets in \cref{fig:noisebudgets}, the discrepancy between the total measured noise and the sum of known noises is most visible below 50 Hz, where the noise from scattered light is typically observed.
The improvements in detector sensitivity have made noise from scattered light more evident at both detectors.
Additionally, with the higher laser powers in O4, shaking tests indicate that scatter couplings have increased at both detectors.

In the commissioning period before O4, both observatories made vacuum incursions to mitigate scatter couplings previously identified during O3~\cite{Soni_2021,O3b}. 
In particular, both detectors removed a septum window at the output port, which separated the volumes housing the signal-recycling mirrors and the output mode cleaner. Previously, the main interferometer beam passed through this window to reach the output mode cleaner, and light scattered from this in-vacuum window coupled excess acoustic noise into the gravitational-wave readout. After removing the septum window, the vibration noise coupling at the output port was significantly reduced, by up to two orders of magnitude in some frequency bands. 
This reduced the total detector noise between \qtyrange{30}{100}{\Hz} at Hanford and between \qtyrange{15}{25}{\Hz} at Livingston, leading to a few megaparsecs of range improvement at both detectors.

Baffles installed throughout the detectors are designed to reduce scattered light by blocking reflection from other surfaces. However, scattering from baffles was an issue identified in O3, where increased vibration levels from e.g, wind and anthropogenic noise produced a harmonic series of noise peaks in the gravitational-wave strain that were characteristic of scattered light modulated by a resonantly shaking scattering surface~\cite{o3paper,O3b}. The scattering sources were identified using targeted vibration injections and movies of scattered light modulation in the test mass chambers. The significant sources were the cryopump manifold baffles and \FP{} arm cavity baffles, both designed to meet scattered light requirements at low motion amplitudes, but whose motions were amplified by high quality-factor resonances~\cite{Soni:2023kqq}. This excess resonant motion produced harmonic series of the resonance frequencies that reached into the sensitive frequency band of the measured strain~\cite{Soni:2023kqq}. 
Mitigation of these baffle scattering sources included damping the baffles or improving their suspensions to reduce the velocity of reflecting surfaces, and shift the frequency of the scattered light noise out of the detection band~\cite{Soni:2023kqq}.

\subsection{Grounding improvements} {\label{sec:grounding_upgrades}}
During the O4 commissioning period, spectra of the variation in current flowing from building electronics ground to neutral earth were observed to be correlated with excess noise in the gravitational-wave strain. A newly developed electronics ground injection system showed that large spectral features in the strain could be produced by injecting 100 milliamp-scale currents onto the building electronics ground. The coupling is thought to be produced by fluctuations in the potential of the electronics ground system due to the variations in current flows across the finite resistance between electronics ground and the neutral earth around the building (measured to be about 2 ohms at Hanford). Forces on the charged test mass may fluctuate with the potentials of nearby electronic systems that are referenced to the fluctuating electronics ground, such as the electrostatic drive and ring heaters. 

The noise from electronics ground potential fluctuations was reduced in two ways. First, the resistance between certain electronics chassis and the building electronics ground was reduced in order to reduce the total resistance to neutral earth for those electronics. Second, the voltage biases of the test mass electrostatic drives were swept and set to values that minimized the coupling of current injections onto the electronics ground to the gravitational-wave strain. It is hypothesized that, at the coupling minimum, the force on the test mass due to its charge is partially cancelled by the force due to the bias-induced polarization. These mitigations improved the range by a few megaparsecs.  
Further mitigation could likely be obtained by shielding electronics inside the chamber from the test mass with shields connected to the chamber.

\section{Quantum noise characterization} \label{sec:characterization}

To reach O4 sensitivities, the quantum-limited detector noise was improved by increasing both the measured squeezing levels and the circulating laser powers. 
This section focuses on the characterization of squeezing and circulating arm power that together determine the shot-noise-limited detector sensitivities, with maximum squeezing levels of up to \SI{\sqzmaxlho}{dB} 
at Hanford and \qty{\sqzmaxllo}{dB} at Livingston observed around 2 kHz during O4b. This level of performance met the O4 squeezing goals for \qty{4.5}{dB} of broadband frequency-dependent squeezing, although stable performance at the maximum squeezing levels remained a challenge. Squeezing performance in O4 has demonstrated that the LIGO detectors are well-prepared to meet the A+ detector goals for \qty{6}{dB} of broadband squeezing.

Meanwhile, both detectors attempted to reach the goals for \qty{400}{kW} of circulating power, instead achieving between \qtyrange{280}{370}{kW} of circulating power. The Hanford observatory was able to briefly observe at \qty{400}{kW} operating power; however, control instabilities and excess noise impacted the detector duty cycle and sensitivity, leading to a power reduction early in O4a. Even with lower-than-targeted circulating powers, the achievements in broadband squeezing have allowed the detectors' strain sensitivities to reach the O4 goals.

\subsection{Squeezing limitations} \label{sec:sqz_characterization}

\begin{figure}[t]
    \centering
    \includegraphics[width=0.48\textwidth,trim={0 0 0 0}]{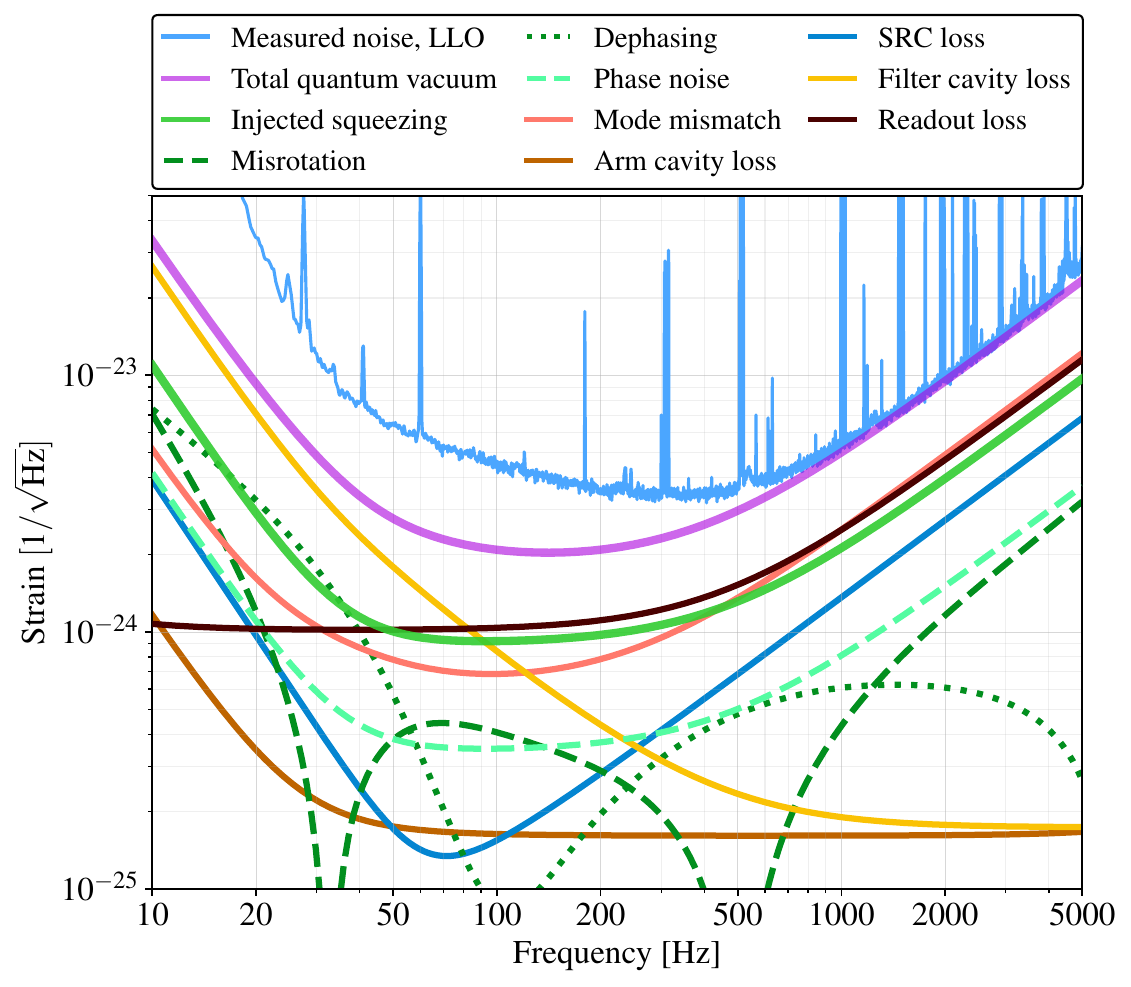}
    \caption{Quantum noise sub-budget with frequency-dependent squeezing injected, shown for the Livingston detector. The total quantum noise (purple) is determined by the circulating arm powers, detector configuration, and squeezer parameters. 
    The injected squeezing level (solid green) is the product of the generated squeezing level and injection losses from the squeezed light source to the output Faraday isolator. At kilohertz frequencies, measured squeezing levels are limited by readout loss (black), interferometer losses (from the arm cavities, brown; and SRC, blue), and mode-mismatch (salmon) between the squeezer, interferometer, and output mode cleaner cavities.  
    At intermediate frequencies, additionally, the detector configuration and detunings~\cite{bncPhysRevD.64.042006}, as well as mode-mismatches~\cite{mccullerPRD21LIGOQuantum}, will introduce squeeze-angle misrotations (dashed green) and frequency-dependent losses that limit the measurable squeezing levels. 
    At low frequencies below \qty{50}{Hz}, squeezing is limited by residual misrotation from the filter cavity (which is not optimally matched to the arm power~\cite{jiasubSQL}), and by filter cavity loss (yellow)~\cite{ganapathyBroadbandQuantumEnhancement2023} that additionally introduces dephasing (dotted green) between the correlated squeezer noise sidebands~\cite{kweeDnD.PhysRevD.90.062006,whittleOptimalDetuningQuantum2020,mccullerPRD21LIGOQuantum}. Squeezer phase noise (dashed light green) between the squeezed beam and local oscillator light (light dashed green) is only a minimal contribution to the total squeezing degradation~\cite{dwyerSqueezedQuadratureFluctuations2013,kijbunchooLowPhaseNoise2020a}.}
    \label{fig:qn_subbudget}
\end{figure}

In O4, the measured squeezing was limited by loss, mode-mismatch, and squeeze-angle misrotation~\cite{Aoki:06,bncPhysRevD.64.042006,mccullerPRD21LIGOQuantum}. \cref{fig:qn_subbudget} shows a sub-budget of the quantum noise with frequency-dependent squeezing for the Livingston detector. 

At kilohertz frequencies, i.e, above the coupled-cavity pole of $\sim$\qty{450}{\Hz}, the shot-noise-limited strain sensitivity is determined by the circulating laser power and the measurable squeezing levels. 
In this region, the usable squeezing is loss-limited: while both detectors expect total optical losses of $\sim$15\%, the measured squeezing implies losses of \qtyrange{20}{30}{\%}, indicating \qtyrange{5}{15}{\%} of unexplained losses. The nonlinear gain measurements at Hanford suggest these excess losses are related to the interferometer readout (see \cref{sec:sqz_level}). However, throughput measurements along the readout path have not found such power discrepancies, suggesting that excess squeezing losses are not simply optical losses, but likely related to alignment and mode-matching. At Livingston, the pre-O4 and mid-O4 analysis further suggest that mode-mismatch and alignment are indeed primary sources of excess loss and squeeze-level drifts over the run. 

Squeezer phase noise, which refers to the phase stability between squeezed light and the local oscillator, negligibly impacted the measured squeezing in O4, as shown by the light green dashed trace in \cref{fig:qn_subbudget}. Squeezer phase noise was estimated to have a root-mean-square total of \qtyrange{20}{27}{\milli\radian} at both detectors~\cite{ganapathyBroadbandQuantumEnhancement2023, jiasubSQL}. Of this, about \qtyrange{5}{10}{mrad} is estimated to be intrinsic to the squeezer control loops, and $\sim$\qty{15}{mrad} was previously estimated to result from the RF sidebands used for interferometer sensing and control~\cite{dwyerSqueezedQuadratureFluctuations2013, kijbunchooLowPhaseNoise2020a}. 

At intermediate frequencies between \qtyrange{100}{450}{\Hz}, where squeezed light begins to resonate in the signal recycling cavity, quantum noise is sensitive to many detector parameters such as the detuning of the signal recycling cavity, the homodyne readout angle, the circulating arm power, and the mode-matching across the detectors~\cite{bncPhysRevD.64.042006,mccullerPRD21LIGOQuantum}. \cref{fig:qn_subbudget} plots the quantum noise model with a direct homodyne readout angle of +11 degrees for Livingston (+10.7 degrees was estimated for Hanford). The magnitude of the readout angle was informed by measurements of the contrast defect power compared to the total output power on the two readout photodiodes. For quantum noise models, the sign of the readout angle was determined from the detector noise spectra with frequency-independent anti-squeezed quantum noise injected between \qtyrange{20}{80}{Hz}.

Below \qty{100}{Hz}, both detectors observe significantly less squeezing. The filter cavity detuning~\cite{kimblePRD01ConversionConventional} must be experimentally adjusted \emph{in-situ} to optimize the quantum noise between \qtyrange{50}{100}{Hz}; this detuning is then estimated from fits to quantum noise models after classical noise is subtracted. Even with classical noise subtraction, almost no squeezing is observed around \qtyrange{30}{40}{\Hz}; at even lower frequencies, both detectors expect that squeezing is largely limited by filter cavities non-idealities. This includes filter cavity losses which are asymmetric between the squeezer noise sidebands, and thus limit squeezing efficiency while introducing sideband dephasing~\cite{kweeDnD.PhysRevD.90.062006,mccullerPRD21LIGOQuantum}. Additionally, the filter cavity linewidth is not yet optimally matched to compensate the arm powers in O4~\cite{whittleOptimalDetuningQuantum2020,jiasubSQL}, leaving residual low-frequency squeeze angle misrotations.

\begin{figure}[t]
    \centering
    \begin{minipage}{0.54\linewidth}
        \centering
        \includegraphics[width=\linewidth,trim={8cm 0 8cm 0}]{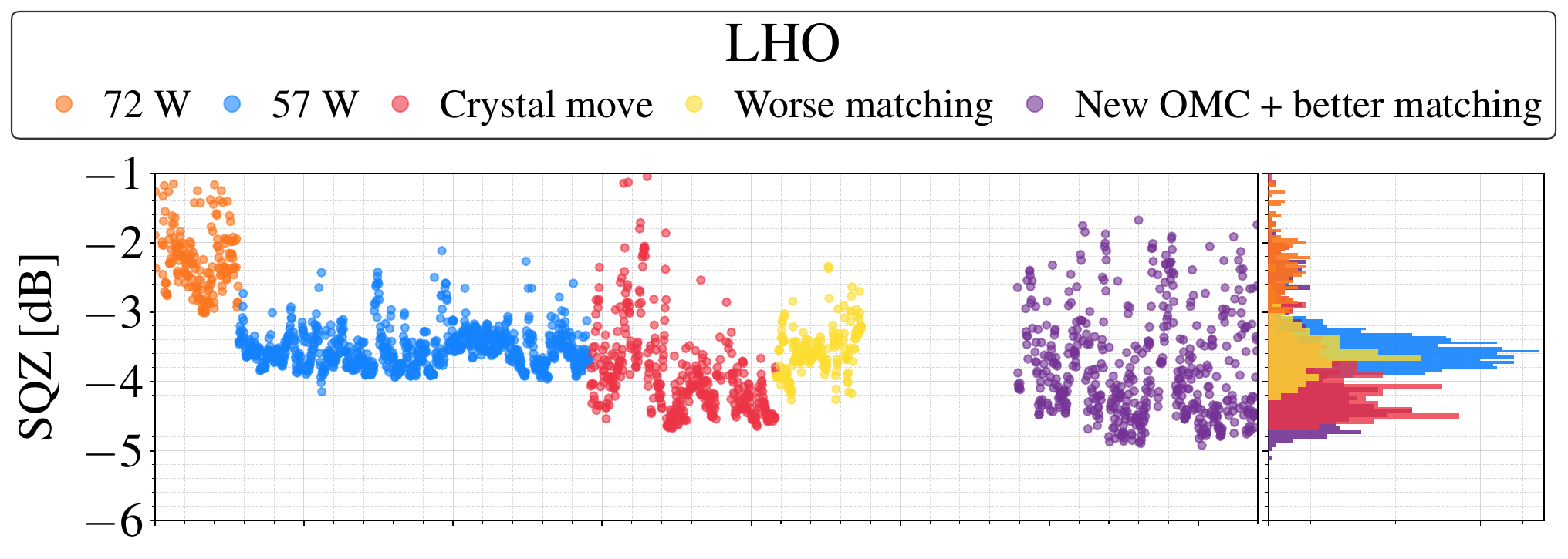}
    \end{minipage}
    \begin{minipage}{0.54\linewidth}
        \centering
        \includegraphics[width=\linewidth,trim={8cm 5mm 8cm 0mm}]{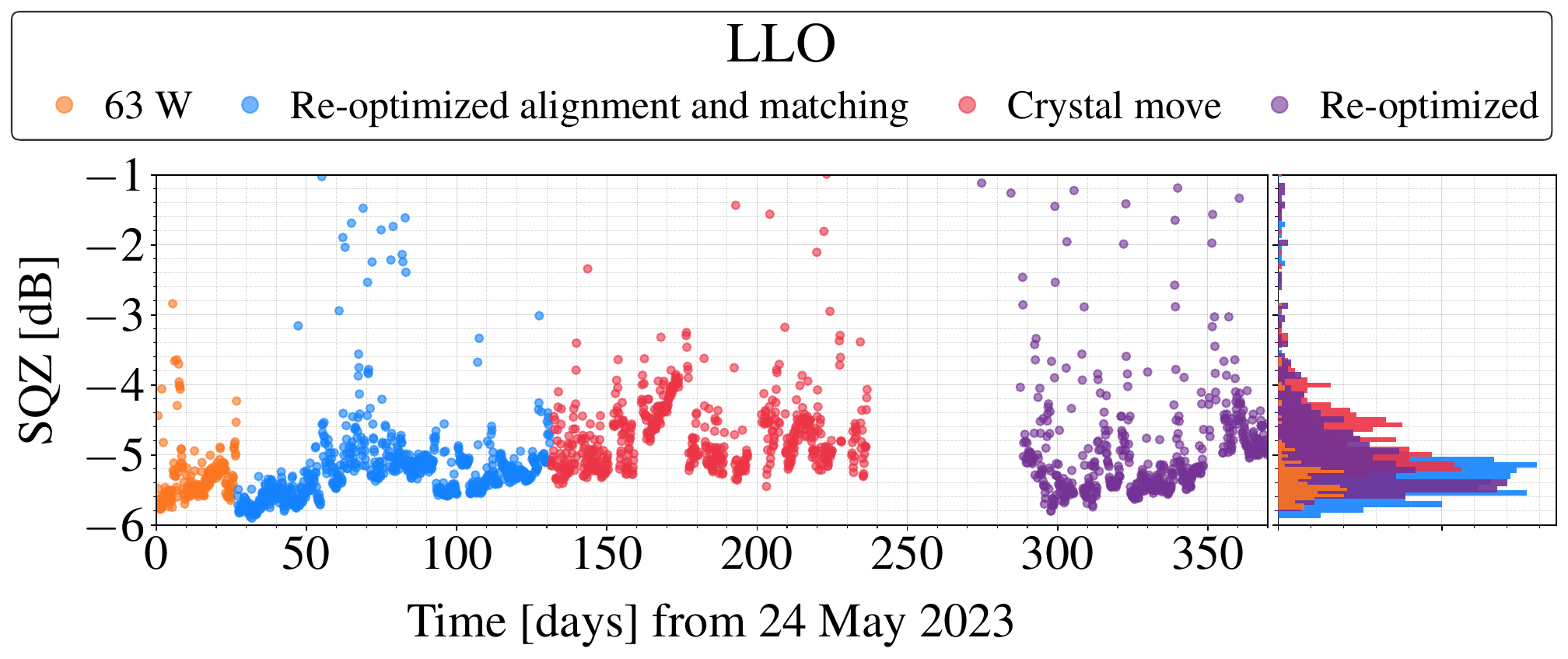}
    \end{minipage}
    \caption{Squeezer performance for the first year of O4 at both Hanford (LHO) and Livingston (LLO) detectors. The vertical axis shows the decibels of total detector noise reduction (negative values) with squeezing injected, compared to unsqueezed shot noise level, as measured around 2 kHz. Colors mark various changes that impacted squeezing efficiency. The noise reduction with squeezing shown here does not include the subtraction of non-quantum noises. }
    \label{fig:sqzDriftsO4}
\end{figure}

\subsection{Squeezing levels over O4} \label{sec:sqz_level}

While squeezing levels are routinely optimized before and during the observing run, the measured squeezing levels varied over the run. 
\cref{fig:sqzDriftsO4} shows the total detector noise reduction in the shot-noise-limited region around 2 kHz when squeezing is injected. The trends in \cref{fig:sqzDriftsO4} thus are not strictly the quantum noise reduction, as classical detector noise is not subtracted in this data. In particular, during high-power operation at the Hanford detector (\cref{fig:sqzDriftsO4}, upper, orange segment), classical noise was estimated to be around 10-fold (in power) below the total unsqueezed shot noise level, implying a higher squeezing level than simply witnessed by the total noise reduction. For the remainder of O4a, at 2 kHz, laser frequency noise was estimated (see Figs. \ref{fig:noisebudgets}, \ref{fig:laser_subbudget}) to be within a factor of 20 (in power, see \cref{fig:corr_noise_budget}) of unsqueezed shot noise, thus, accurately assessing the efficiency of squeezing for quantum noise reduction requires the subtraction of classical (i.e, non-quantum) detector noise. For the Livingston detector, the classical noise level around 2 kHz is almost 100-fold in power below unsqueezed shot noise, allowing these trends to be a more accurate witness of the detected squeezing level.

\begin{figure}[t]
    \centering
    \includegraphics[width=0.48\textwidth,trim={0 5mm 0 0mm}]{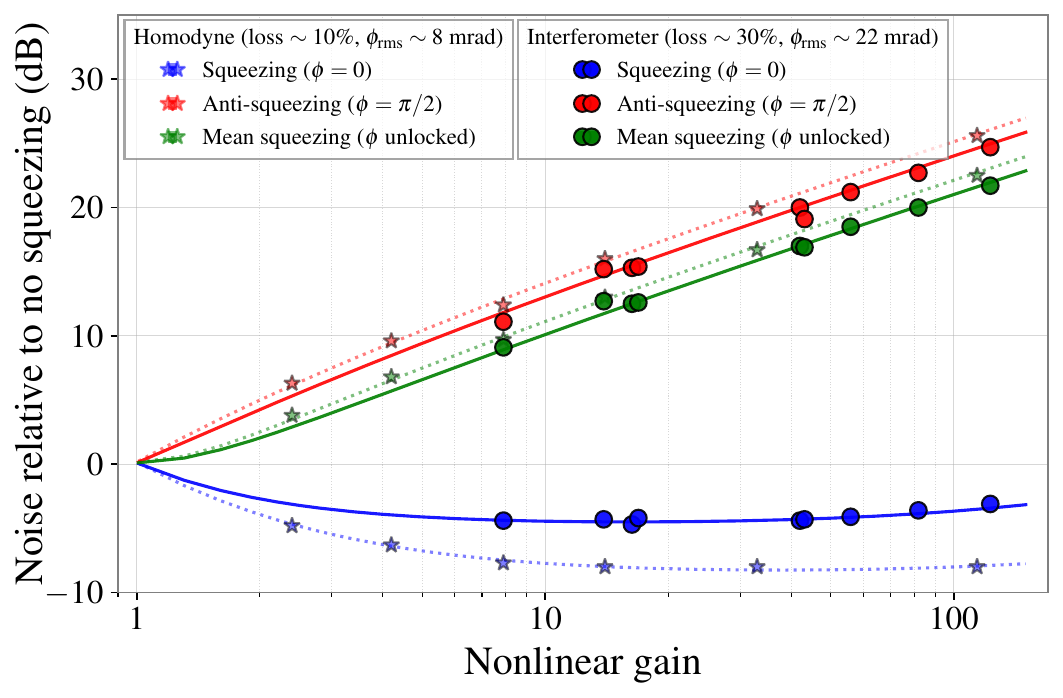}
    \caption{Squeezing levels measured as a function of generated squeezing levels at LIGO Hanford. Squeezing is measured using i) a diagnostic in-air homodyne detector before the interferometer (stars, dotted lines), and ii) the GW readout photodiodes at 2 kHz, after propagation through the full interferometer (circles, solid lines). Lines show fits to the data, with the fitted loss and phase noise shown in the legend. For the homodyne, 8 dB of measured squeezing is consistent with known in-chamber squeezing injection loss and table-top readout loss, leaving no unresolved loss. For the interferometer, the measured squeezing levels suggest $\sim$30\% loss and $\sim$20 mrad of phase noise, exceeding the expected squeezing losses of 15-20\%.
    }
    \label{fig:h1_nlg_sweeps}
\end{figure}

\emph{Squeezing loss characterization.}
At Hanford, squeezing levels were measured at a function of nonlinear gain (i.e, as a function of generated squeezing levels) to experimentally constrain squeezing loss and phase noise with the full detector, as shown in \cref{fig:h1_nlg_sweeps}.
Squeezing levels were measured around \qty{1}{kHz} using (i) a diagnostic in-air homodyne detector (stars, dotted line) that beats the squeezed beam before injection into the interferometer (just before the output Faraday isolator) against local oscillator light from the main laser, and (ii) the interferometer's output photodetectors (circles, solid line), when the full interferometer stably locked and thermalized. 
The homodyne measurement was used to validate the expected injection losses of about 7.5\%, which include known losses from the squeezed vacuum source to the interferometer's output faraday isolator. 
In this measurement, the measured homodyne squeezing levels before the interferometer were well-explained by the known injection losses, thus leaving no unexplained squeezing losses before the interferometer. 
Interferometer squeezing measurements could then be used to constrain the detector's total squeezing loss and phase noise.

At Livingston, extensive quantum noise models and parameter estimation of detector parameters, with priors informed by measurements, were used to understand squeezing limitations. A quantum noise model was initially established using Markov-Chain-Monte-Carlo (MCMC) analysis of the detector quantum noise for an extensive dataset taken directly before the start of O4~\cite{jiasubSQL}.
A second MCMC analysis was performed mid-run with a shorter data stretch, using the pre-O4 parameters~\cite{jiasubSQL} as priors. This mid-run analysis provided the quantum noise estimate in the Livingston noise budget of \cref{fig:noisebudgets}. 
Comparison of the pre-O4 and mid-O4 analysis suggests that, after several months of observing, the squeeze level drifts seen in \cref{fig:sqzDriftsO4} (lower) largely correspond to mode-matching degradations. After several months of observing, the inferred mismatch between the interferometer and OMC was about 6\% (i.e, 2\% worse than the start of the run), and the inferred mismatch between the squeezer and OMC was 4\% (i.e, 3\% worse). Correspondingly, fine-tuning and optimization of the squeezer alignment and mode-matching in O4b successfully improved the measurable squeezing levels to a maximum of \qty{\sqzmaxllo}{dB}.

\emph{Alignment control.}
To reach the maximum squeezing levels for O4, the Livingston detector discontinued use of the active squeezer alignment controls previously used in O3~\cite{tsePRL19QuantumEnhancedAdvanced}. Instead, when the detector alignment was impacted, alignment optimization scans were used; these routines iteratively walked two mirrors along the squeezing injection path in order to maximize the measured squeezing levels around 2 kHz. The Hanford detector used active alignment control in O4a (as described in ~\cite{ganapathyBroadbandQuantumEnhancement2023}), and after a brief period of discontinued use for the first few months of O4b, returned to using alignment control for the latter part of O4b. The signal-to-noise ratio used for the squeezer alignment controls signals is low, and cross-couplings are observed between the injected squeeze angle and squeezer misalignment, complicating the use of squeezer alignment control to reach the maximum observable squeezing levels. At Hanford, the best performance was observed when using squeezer alignment control alongside squeeze angle control, which controls the injected squeeze angle based on an audio-band diagnostic field~\cite{adf.PhysRevD.105.122005} that senses the detected output squeeze angle.

\emph{Mode-matching.}
To optimize mode-matching between the squeezed beam, interferometer beam, and output mode cleaner, the active mode-matching optics were varied to maximize measured squeezing levels. At Hanford, these investigations led to a vacuum incursion between O4a and O4b to extend the tunable range of the PSAMS optics for further optimization; this incursion indeed enabled higher squeezing levels in O4b, as seen in Fig. \ref{fig:sqzDriftsO4} (upper, yellow/purple dots). 
At Livingston, mode-matching scans were performed after major detector configuration changes (e.g, input power or thermal state changes). These mode-matching optimization routines consisted of iteratively adjusting the radii of curvature for the two PSAMS mirrors on the squeezing injection path, in order to maximize the observed squeezing levels at kilohertz frequencies.

\emph{Squeezer crystal.}
Issues with squeezer crystal degradation persisted during O4. The squeezed light source is a dually-resonant bowtie cavity containing a nonlinear crystal of periodically-poled potassium titanyl phosphate (PPKTP)~\cite{tsePRL19QuantumEnhancedAdvanced}. 
In October 2023, both sites translated the PPKTP crystal within the squeezer cavity to reduce green crystal absorption and losses, with additional motivations at Hanford to recover squeezing losses. 
\cref{fig:sqzDriftsO4} shows the effects of translating the crystal for both detectors. The Hanford detector recovered about 8\% of squeezing loss, reaching higher squeezing levels and higher detector sensitivities after crystal translation (upper, red). This improvement corresponded to the detector sensitivity improvement in October 2023, which can be seen in \cref{fig:rangeplots}. Meanwhile, the crystal translation at the Livingston detector (lower, red) slightly lowered the generated (inferred) squeezing level from about \qty{17.4}{dB} to \qty{11.5}{dB}, but did not otherwise correspond to major changes in squeezing levels or detector sensitivity. Both detectors translated their squeezer crystals again in O4b, with minimal impact on the measured squeezing levels.

\subsection{Arm power}
\label{sec:arm_power_characterization}

Estimating the circulating power in the \FP arm cavities is another important aspect of quantum noise characterization.
As described in \cite{o3paper}, the arm power is not trivial to constrain due to uncertainties in the incident power on the beamsplitter, the arm cavity gain, and the readout losses. This section describes some of the methods and challenges of quantifying the arm power, as the LIGO detectors reach their highest circulating powers to date.

One of the simplest ways to estimate arm power uses the measured input power and power-recycling cavity finesse. Specifically, this considers the input power incident on the power-recycling mirror ($P_\text{in}$), the power gain from the finesse of the power-recycling cavity (i.e, the power-recycling gain, $G_\text{PR}$), and the arm cavity gain ($G_\text{arm}$) given the input and end cavity mirror transmissivities and the round-trip optical losses. $G_\text{PR}$ relates the power incident on the main beamsplitter ($P_\text{BS}$) to the power on the power-recycling mirror ($P_\text{in}$): $G_\text{PR} = P_\text{BS}/P_\text{in}$. The circulating arm powers can be then calculated as
\begin{equation}
    P_\text{arm} = \frac{1}{2} P_\text{in} G_\text{PR} G_\text{arm}.
    \label{eq:arm_power}
\end{equation}
This method was previously applied to estimate the arm cavity power in O3~\cite{o3paper}. 

This estimate is made more accurate by estimating what fraction of the input power, $P_\text{in}$, is in the fundamental TEM$_{00}$ mode and at the carrier frequency (i.e, not RF sideband power). For both detectors, a few percent of the total input power is measured to be in higher order transverse spatial modes, and at RF sideband frequencies as needed for interferometer control. Higher order mode content of the input beam is constrained by measuring the ratio of transmitted to reflected carrier power, as most of the power reflected off the interferometer will be due to mode mismatch in the input beam. RF sideband content is inferred by adjusting the modulation depth of the RF sidebands, and measuring the resulting change in power at the symmetric and anti-symmetric ports. The main sources of uncertainty in this circulating power estimate stem from calibration of the photodetectors used to estimate the input power, RF sideband power, and the higher order mode content within the input beam. 

\begin{table*}[t]
\setlength{\tabcolsep}{10pt}
\centering
\caption{Summary of the operating power parameters of the LIGO Hanford and Livingston interferometers during O4a, using multiple estimation methods. The uncertainties in arm power calculated from \cref{eq:arm_power} result from the propagation of the photodiode uncertainty used to estimate input power and cavity gains. The uncertainties in the estimation of arm power from the quantum noise result from calibration uncertainties, non-stationary noise, low measurement averaging time, and model degeneracies.
} \label{tab:power_params}
\begin{tabular}{l c c c} 
    \hline \hline  Parameter & Hanford, O4a & Livingston, O4a & Units \\
    \hline
    Input laser power 
     & 57 $\pm \ 2 $ & 64 $\pm \ 1 $ & W  \\
    Input carrier 00 power,  $P_{\mathrm{in}}$
     & 56 $\pm \ 2 $ & 62 $\pm \ 2 $ & W  \\
    Power-recycling gain, $G_{\mathrm{PR}}$
       & 50 $\pm \ 1$  & 35 $\pm \ 1$ & W/W \\
    Arm cavity power via \cref{eq:arm_power}, $P_{\mathrm{arm}}$   
       & 364 $\pm \ 15$ &  285 $ \pm \ 10$  & kW  \\
    Arm cavity power via quantum estimate & 360 $ \pm \ 10$ & 260 $ \pm \ 5 $ & kW \\
    \hline
    
\end{tabular}
\end{table*}

The corresponding values of each of these parameters for both detectors are listed in \cref{tab:power_params}.
Previously in O3, the arm power at Hanford was estimated to be around 195$\pm$14 kW, as reported in Ref.~\cite{o3paper}. The replacement of the Y-arm input test mass at Hanford (see \cref{sec:testmasses}) corresponded with a significant reduction in the arm cavity losses, and a corresponding increase in the recycling gain at high power. The resulting power-recycling gain is estimated to be 50.1 W/W at 57.5 W input power. At this power, 97.2\% of the input laser is power is carrier power, and the rest is higher order mode and RF sideband power. The total uncertainty on the estimate of input carrier 00 mode power is 2.8\%.
Applying the measured arm cavity gain of 260, the resulting arm power estimate is 364$\pm$15 kW following \cref{eq:arm_power}. This estimate assumes the cavity gain for each arm is the same.

The Hanford detector operated at a higher input power of \qty{72}{W} during the first month of O4a. 
At \qty{72}{W} input power, output losses increased slightly due to increased thermal distortions which led to greater mode mismatch within the interferometer. This mode mismatch reduced the estimated power-recycling gain to 45.6 W/W at 72 W input power. An estimated 4.4\% of the input power was estimated to be in higher order modes and RF sidebands due to this excess mode mismatch. Using the same arm cavity gain of 260, the resulting arm power estimate was 407$\pm$16 kW. Section~\ref{sec:highpower} discusses the power reduction to 57 W during the early part of the observing run.

Previously in O3, the arm power at Livingston was estimated to be around \qty{240(18)}{\kW}, as reported in Ref.~\cite{o3paper}. For O4, the replacement of the end test mass mirrors at Livingston corresponded to a small decrease in the power-recycling gain due to an increase in the arm cavity losses from higher scattering and absorption (see \cref{sec:testmasses}). The resulting power-recycling gain was calibrated to be 34.8 W/W at 64.5 W input power. Within the input beam, 4\% of the total power was estimated to be in higher order modes and RF sidebands. The total estimated uncertainty on the input power in the fundamental carrier mode is 2\%.
Applying the measured arm cavity gain of 265~\cite{o3paper}, the resulting arm power estimate was \qty{285(10)}{\kW} following \cref{eq:arm_power}. Similar to Hanford, this estimate assumes the cavity gain for each arm is the same. These parameters are also listed in \cref{tab:power_params}.

\begin{figure}[t!]
    \centering
    \includegraphics[width=0.48\textwidth,trim={0 0 0 0}]{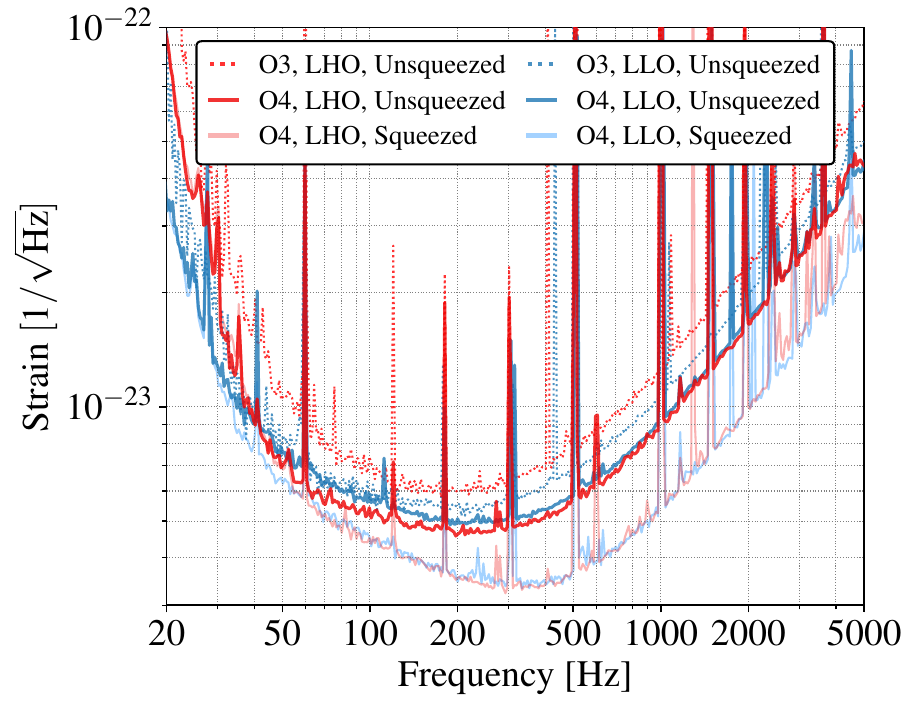}
    \caption{Comparison of the unsqueezed detector noise between O3~\cite{o3paper} (dotted lines) and O4 (dark solid lines). Without squeezing, the shot-noise-limited sensitivities at high frequencies above 500 Hz depends primarily on the circulating arm powers and output losses. With squeezing injected (light solid lines), the two detectors can reach comparable sensitivity.
    }
    \label{fig:compareUnsqzDARM}
\end{figure}

Quantum noise and squeezing measurements can also help constrain the arm power.
Without squeezing, the arm power can be estimated from the calibrated detector strain noise at shot-noise-limited frequencies above $\sim$\qty{450}{Hz}, where the strain noise power scales inversely with the arm power. 
The conversion from the output photocurrent (in milliamps) to the differential arm length change (in meters) is often referred to as the ``optical gain'' for the strain signal; the detector's optical gain is measured in the regular calibration sweeps~\cite{Sun:2021qcg,O4cal}. 
Below $\sim$\qty{100}{Hz}, quantum radiation pressure noise power scales proportionally with the arm power; without squeezing, it is estimated to be almost 10-fold lower in power than the total noise at \SI{20}{Hz}, making this difficult to use for arm power estimates. 

\cref{fig:compareUnsqzDARM} compares the unsqueezed detector noise between O3 and O4. As the expected readout losses are similar for the two detectors, the greater increase in unsqueezed sensitivity reflects the greater increase in circulating power and optical gain for the Hanford detector from O3 to O4.

With injected squeezing, the total measured noise depends strongly on the injected squeeze angle. Fitting detector noise spectra measured at various squeezing angles to a common detector quantum noise model can thus provide a strong constraint on the arm power, as previously described in Refs.~\cite{mccullerPRD21LIGOQuantum, jiasubSQL}. Still, uncertainties in the arm power remain~\cite{yuN20QuantumCorrelations,jiasubSQL} due to model degeneracies between e.g, arm power, readout angle, signal recycling cavity detuning, and readout losses, and due to measurement uncertainties such as calibration uncertainty~\cite{Sun:2021qcg,O4cal}, averaging time especially for low-frequency noise, and non-stationary noise~\cite{Soni_2021}. 

At the Hanford detector, arm power measurements based on quantum noise, and based on the power-recycling gain and input power (as in \cref{eq:arm_power}), are both consistent with arm powers in the range of \qtyrange{350}{370}{\kW}. 
At the Livingston detector, quantum noise measurements with squeezing strongly suggested an arm power closer to \qty{260}{\kW}~\cite{jiasubSQL}, which is lower than the estimate from the power-recycling gain and input power using \cref{eq:arm_power}.

\section{High power challenges}\label{sec:highpower}

In O4, Both the Hanford and Livingston detectors achieved higher circulating laser powers in order to increase their quantum-limited strain sensitivities. 
However, the power increase at both observatories came with several challenges related to thermal compensation and thermalization (\cref{sec:tcs} and \ref{sec:thermalization}), parametric instabilities (\cref{sec:PImitigation}), controls instabilities, and excess low frequency noise. In particular, the Hanford observatory reduced its operating power one month after the start of O4a in order to improve the duty cycle and sensitivity, as described \cref{sec:powerreduction}.

These power increases were enabled by upgrades to the pre-stabilized laser system for higher input powers (\cref{sec:psl}) and the replacement of test masses (\cref{sec:testmasses}) to remove the point absorbers that limited further power increases previously in O3~\cite{o3paper, brooksPointAbsorbersAdvanced2021a}. 
In addition, saturations of the power-recycling cavity pick-off photodetector (POP, see \cref{fig:ligo}) initially prevented increasing the input power above 55 W; this photodetector is the main sensor for the three auxiliary length degrees of freedom: the power- and signal-recycling cavity lengths and Michelson cavity length.
Prior to O4, both detectors performed a vacuum incursion to attenuate the beam along this path to avoid photodetector saturation and allow power up efforts to proceed. This attenuated light along both the POP path and the reflection port path (REFL) used to control the common arm length and the common alignment controls (see \cref{fig:ligo}).

Overall, increasing the operating power at both detectors was hampered by significant challenges related to mode matching, controls, and excess noise. Even when higher power was achieved in the LIGO detectors, the full benefits were not necessarily evident due to the accompanying increased losses. This section summarizes the various challenges of increasing the operating power at both observatories. 
Since advanced gravitational wave interferometers must operate with significant intracavity power, the results presented in this section provide an important accounting of the impacts of high-power operation, some methods to achieve stable detector operations at high power, and the shortcomings of the current compensation systems.

\subsection{Thermal compensation} \label{sec:tcs}

The thermal compensation system (TCS) is designed to sense and correct the thermal lenses and aberrations on the core optics that occur due to test mass coating absorption \cite{Brooks:16}. At both detectors, the uniform test mass absorption is estimated to be around 0.5 ppm, dominated by the optical absorption of the mirror coatings.
At O4 operating powers, the mirrors absorb hundreds of milliWatts of power, resulting in an induced thermal lens that distorts the finely tuned cavity dynamics. This estimated uniform absorption is further impacted by any point defects on the test mass mirrors~\cite{brooksPointAbsorbersAdvanced2021a}. The effect of these point absorbers was discussed at length in Ref.~\cite{o3paper}.

Two main actuators provide thermal compensation to the core optics, as described more fully in Ref.~\cite{Brooks:16}: ring heaters placed around the barrel of each test mass mirror optic, and CO$_2$ lasers incident on a compensation plate behind each input test mass. The ring heaters are designed to create a negative thermal lens on the test masses that corrects aberrations on the highly-reflective surface of the mirrors. The CO$_2$ lasers can induce a positive or negative thermal lens on the compensation plate. An annular beam mask can be used to closely mimic the effect of the ring heaters, while a central beam mask can be used to mimic the heat load of the interferometer beam.
Both detectors tuned their TCS settings, especially using the ring heaters, to improve mode-matching and optical gain, reduce noise couplings to laser frequency and beam jitter noise, and to mitigate parametric instabilities. 

\emph{Hanford TCS tuning.}
With the Hanford detector at \qty{57}{W} input power, the ETM ring heaters were adjusted in both common and differential steps to provide better arm mode matching and reduce contrast-defect light measured at the output port. This had the further benefit of reducing the frequency noise coupling. By increasing the end test mass ring heater power in a common step, the frequency noise coupling was reduced by a factor of three and the contrast-defect light by a factor of five. This change led to the excitation of a parametric instability around 10.4 kHz on ETMY which required active electrostatic damping, as further described in \cref{sec:PImitigation}.

Meanwhile, slight differential tuning of the ETM ring heaters reduced the contrast defect (i.e, improved the optical gain) further, but excited a different parametric instability near 80.3 kHz on ETMX. At the time, no active damping solution was found, and so ETM ring heaters were instead adjusted to mitigate the instability, leading to a slightly higher contrast defect.

To further compensate the increased thermal lens at \qty{57}{W}, common and differential changes of the ITM ring heaters were tested.
A small adjustment of the ITMX ring heater to \qty{0.44}{W} was made to compensate for the thermal aberration due to a point absorber on the test mass.
However, no other adjustments of the ITM ring heaters led to successful operation; the increased negative lens on the ITMs caused instabilities in the controls at low power that precluded lock acquisition. Engaging the ring heaters involves a long transient that can take more than 12 hours to settle~\cite{Wang:2016wsb,Hardwick:2020ukv}, so any ITM ring heater solution found to improve optical gain at full operating power must also permit locking at low power. Because the RF sidebands are antiresonant within the \FP arm cavities, perturbations in the ITM lens have a significant impact on their overlap with the carrier beam. This effect is also at play at full operating power, where the induced thermal lens on the ITMs has a more significant effect on the RF sidebands than the carrier beam. As such, the Hanford observatory operates at high power without full compensation of the ITM thermal lensing, leading to increased control challenges and losses.

Annular CO$_2$ heating is also used while at high power operation to compensate for the thermal load from the interferometer beam. Annular CO$_2$ power was adjusted to reduce noise couplings to laser frequency and beam jitter noise, using similar common and differential step tests.  At \qty{57}{W}, both CO$_2$ lasers are set to apply \qty{1.7}{W} of power to the ITMs using an annular mask. The injected CO$_2$ power is turned to zero on each interferometer lockloss to follow the reduced thermal load of the interferometer beam at lock acquisition.

After the input power increase to \qty{72}{W} at Hanford, similar
common and differential steps of the ETM ring heaters were attempted to improve optical gain and reduce noise couplings. However, the contrast defect at this power increased by 50\% (leading to worse optical gain), and the frequency noise coupling increased due to the increased arm cavity mismatch. Similar thermal tuning methods applied at \qty{57}{W} could not compensate for these losses. Again, all tests of the ITM ring heaters were unsuccessful at improving mode matching and continued to severely affect locking procedures. When the power was reduced to \qty{57}{W}, the ETM ring heaters were reverted to the \qty{57}{W} settings described above.

\emph{Livingston TCS tuning.}
At Livingston, ETM ring heaters were first used to avoid driving parametric instabilities, particularly around \qty{10}{kHz} and \qty{80}{kHz}. Then, differential adjustments of the ITM ring heaters were used to reduce coherence to laser frequency noise around \qty{3}{kHz}. Next, common tuning of the ITM ring heaters targeted improving the build--up of the RF sidebands in the auxiliary length degrees of freedom. After these adjustments on the ITM ring heaters, the ETM ring heaters were adjusted to maintain the higher order mode spacing in the cavity.

Before O4a, increasing the input power at the Livingston observatory did not result in a proportional increase in sensitivity due to loss of mode matching from the increased power. Increasing the ring heater power of both the input and end test mass ring heaters recovered some of the optical gain lost with each power increase, but not all.
With the reduction in losses from the ETM cleaning (see \cref{sec:testmasses}) during the O4 break, further optical gain was recovered for the same operating power compared to O4a. However, Livingston continued to operate at \qty{64}{W} in O4b due to control challenges at higher power. Similar to the Hanford observatory, necessary increases in the ring heater power on the ITMs for improved mode--matching impacted locking procedures, also preventing operation at higher power.

\subsection{Parametric instability mitigation} \label{sec:PImitigation}
Parametric instabilities (PI) are caused by the optomechanical interaction between the radiation pressure in higher--order transverse modes of the arm cavities, and the bulk mechanical resonant modes of test mass mirrors~\cite{EVANS2010665}. 
This interaction can become an unstable positive feedback loop that prevents high--power interferometer operation. 
In particular, higher arm powers result in more absorbed power and thus greater deformation of the test mass optics. This absorption changes the cavity geometry such that more power may be scattered into higher-order optical modes, and the higher--order optical mode frequencies can overlap to excite mechanical PI modes. Thus, PI mitigation becomes more challenging at higher powers.

Such PI modes are observed at both detectors \cite{PhysRevLett.114.161102}, for example as the narrow lines visible in the output photodetector spectrum in the bottom panel of \cref{fig:thermalization}. Both passive damping using piezo-electric acoustic mode dampers (AMDs)~\cite{biscansAMDs} and active damping using the electrostatic drive to the test masses~\cite{PhysRevLett.118.151102} are used to mitigate PIs. AMDs were first installed on the test masses prior to O3 to damp the Q-factor of known mechanical resonances between \qtyrange{15}{80}{\kHz}~\cite{biscansAMDs}.

With the arm power increase for O4, a PI at 10.4 kHz was observed at Hanford, while a PI just above 80 kHz was observed at both Hanford and Livingston. Throughout O4, the Hanford detector used the electrostatic drive to ETMY to actively damp a 10.4 kHz PI, which was optically driven by the circulating power in the second-order Hermite-Gauss transverse mode ($10.4$ kHz $\approx 2\times5.2$ kHz~\cite{o3paper}). This PI was excited due to the intentional beam mis-centering to avoid point absorbers on the Y-arm mirrors (see \cref{sec:testmasses}). 

Both detectors tuned ring heater settings~\cite{Brooks:16} to mitigate a PI around 80 kHz, which was likely excited by the optical power in a first higher-order transverse mode that is two free spectral ranges ($\mathrm{FSR}\sim37.5$ kHz) away from the carrier (i.e, $5.2\ \text{kHz} + 2\times37.5\ \text{kHz} \approx 80$ kHz).
The upgrade to a \qty{524}{kHz} ADC, detailed in \cref{sec:darknoise}, enabled sensing the \qty{80}{kHz PI} in transmission of the output mode cleaner.
At Hanford, a 80.3 kHz PI was observed on ETMX at \qty{57}{W} input power, and was mitigated by tuning the ETMX ring heater to offset the optical mode from the mechanical mode. At Livingston, a 80.4 kHz PI appeared on ITMY at 50 W input power, and was mitigated by tuning the ITMY and ETMY ring heaters.

\subsection{Thermalization}\label{sec:thermalization}

\begin{figure*}[t!]
    \centering
    \includegraphics[width=0.9\textwidth]{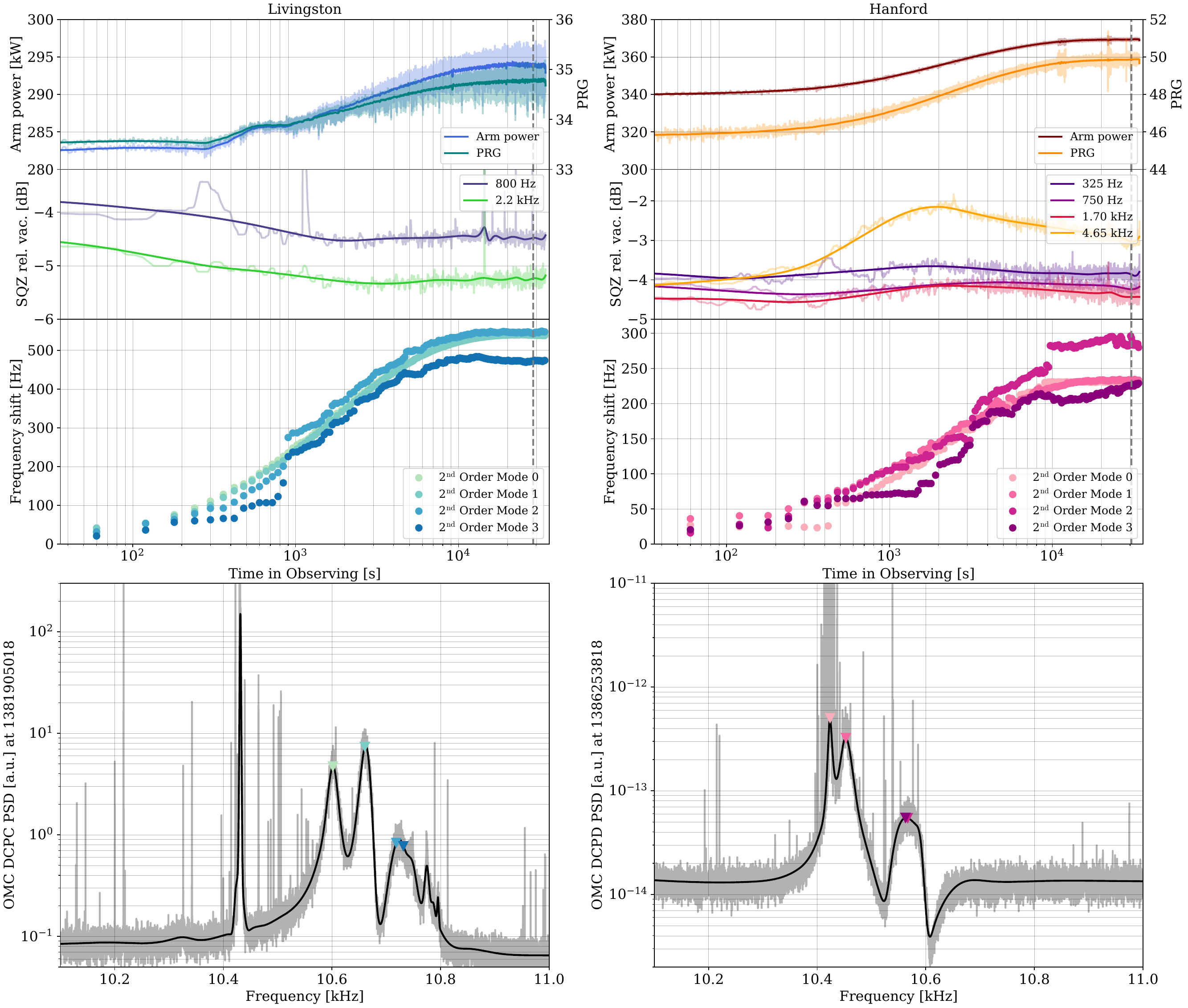}
    \caption{Thermalization trends at the LIGO Livingston Observatory (left) and LIGO Hanford Observatory (right), for the O4a lock stretches in which the detector noise spectra of \cref{fig:noisebudgets} is measured. Upper panels show i) the increase of circulating arm power and power recycling gain, ii) the recovering of squeezing levels during thermalization, and iii) the drift of the second transverse mode frequency, showing how the cavity geometry drifts as the test mass mirrors thermalize and deform from the optical power absorbed in the mirror coatings. 
    The bottom panel shows the output photodetector sum signal around the second transverse mode frequency near 10.4 kHz at the time indicated by the grey dotted vertical lines on the upper plots. The broad peaks are visible due to the noise coupled in by mode-mismatch around the second higher-order transverse mode frequencies; plots tracking the frequency drifts of these second-higher order mode are shown just above in iii). The sharp peaks largely correspond to the bulk mechanical modes of the test mass mirrors, which are susceptible to the opto-mechanical parametric instabilities described in \cref{sec:PImitigation}.
    }
    \label{fig:thermalization}
\end{figure*}

Both detectors experienced challenges related to \textit{thermalization}: transients in the operating point of the detector due to the optical power absorbed from the interferometer beam into the test mass mirror coatings, and the subsequent evolution of the induced thermal lens.  
Previous demonstrations of dynamic thermal compensation revealed that this thermal evolution can last for several hours at the beginning of each lock as the test masses absorb hundreds of mW of power~\cite{Hardwick:2020ukv}. 
The thermal compensation system is designed to counteract induced thermal lensing in the steady-state~\cite{Brooks:16}, but is less effective at dynamically counteracting the changing thermal lens that occurs at the beginning of each lock.

During thermalization, beam parameters of the main beam and RF sidebands change dramatically, higher order mode frequencies shift, and noise couplings change as the mirrors absorb power. The effects of thermalization can limit the operating power of both detectors; it is challenging to compensate the thermal lens generally, and more challenging to compensate the early thermal evolution each lock. \cref{fig:thermalization} shows an example of interferometer signal trends during a lock sequence. Arm power steadily increases for the first three hours of the lock as the power-recycling gain increases. Following a similar evolution, squeezing levels and higher order mode frequencies shift. The shifting of the higher order mode frequencies indicates the changing cavity geometry as a result of the induced thermal lens from the central heating of the interferometer beam. The evolution of squeezing levels results from the change in the  mode-matching between the interferometer and squeezer.

For both detectors, in addition to its effect on the optical gain and noise couplings, the thermalization process impacted the measured squeezing. As the inteferometer thermalizes, the optimal squeezing angle, squeezer alignment, and squeezer mode-matching can all change. 
The squeezer optimizations are made with the interferometer after it is thermalized. Therefore, squeezing levels are not optimal during thermalization, and the maximum squeezing levels are typically reached only 3-4 hours after the interferometer reaches full power, as seen in the second panel of \cref{fig:thermalization}. 
At the Hanford detector, the detuning of the signal-recycling cavity is also observed to drift during thermalization, leading to further frequency-dependent drifts of the measured squeezing~\cite{mccullerPRD21LIGOQuantum}.

\subsection{Hanford power reduction} \label{sec:powerreduction}

After the first month of observing in O4a, the Hanford detector decreased the input power from \qty{72}{W} to \qty{57}{W}, reducing the estimated circulating power from around 407 kW to 364 kW as listed in \cref{tab:power_params}.  
Operating at \qty{72}{W} input power came with significant controls challenges related to thermalization, which severely impacted the duty cycle and sensitivity of the Hanford detector. 

While the carrier beam power increased during thermalization, the sideband gain decreased dramatically which deeply affected the lock stability and noise couplings, following the evolution of the uncompensated ITM thermal lens. In particular, the 9 MHz sideband used to control the power-recycling cavity length lost up to 75\% of its optical gain during the first few hours of every lock.

A servo to maintain a stable unity gain frequency was commissioned to compensate this loss, but the gain evolution was unpredictable in each lock, as the overall thermal state of the interferometer varied significantly from lock to lock depending on how quickly the interferometer could reacquire lock to the high power thermal state.

This control instability of the 9 MHz sideband caused regular fast locklosses within the first few hours of observing. The reduced duty cycle of the Hanford detector in O4a can largely be attributed to this period of high power operation; the power reduction improved the stability of the auxiliary controls and in turn the detector duty cycle.

The significant evolution of the sidebands further impacted the ability to control noise couplings consistently during each lock. With significant evolution in the sideband gain used to control the auxiliary length degrees of freedom, the length feedforward tuning described in \cref{sec:lengthcontrolnoise} became a challenge. A feedforward scheme measured at the start of a lock would lose efficacy during thermalization, leading to increased low frequency noise during observing periods. 
At \qty{72}{W} input power, thermalization was also observed to have a more severe effect on squeezing, which further degraded detector sensitivity over the first few hours of each lock.

While operation at lower power still included a thermal transient for the first few hours of the lock, the effects of thermalization reduced significantly at \qty{57}{W} input. 
In addition, decreasing the input power from 72 to \qty{57}{W} directly corresponded to a decrease in broadband noise between \qtyrange{20}{50}{\Hz}. Some of the elevated noise at high power could be explained by excess noise in the length control loops, which, as described above, were not well-mitigated by the length feedforward. However, even after coherent subtraction of the noise associated with the length controls, there was an approximate doubling of the broadband noise at low frequency.  Most of this noise could likely be attributed to vibrational noise causing excess scattered light noise and beam jitter noise, which is not well-witnessed enough for coherent subtraction.
Some of this excess can result from the imperfect compensation of the thermal distortions which cause the \FP{} arms to become more mismatched, coupling more jitter noise to the gravitational-wave readout due to higher contrast defect. Mirror defects, such as the point absorbers discussed in \cref{sec:testmasses}, absorb more power and cause more uneven mirror distortion, increasing the fraction of scattered light and therefore the coupling of vibrational noise.
The reduction of low frequency noise with lower power allowed further noise and squeezing improvements to increase detector sensitivity throughout the run, as shown in \cref{fig:rangeplots}.

\section{Future work} \label{sec:future}

After O4, a break will precede the fifth observing run (O5) to commission major detector upgrades, continuing the implementation of the A+ upgrade program~\cite{abbottProspectsObservingLocalizing2020}. 
To reduce coating thermal noise (\cref{sec:ctn}), which becomes a more dominant noise source as the quantum noise is reduced, new test mass mirrors with upgraded coatings will be installed~\cite{PhysRevLett.127.071101}. 
Replacing all test masses may also resolve remaining issues of point absorbers which, as discussed in \cref{sec:testmasses}, could further improve control stability and noise couplings to ultimately enable higher power operation. 
The A+ upgrade plan also involves installing a larger beamsplitter to reduce clipping losses in the detectors. 
To mitigate input beam jitter noise, which in O4 degraded the sensitivity of both detectors in multiple bands between 100--1000 Hz (\cref{sec:beamjitter}), a future upgrade involves adding a new pre-mode cleaner between the laser bench and the input mode cleaner to filter out beam jitter~\cite{JAC}.

To increase the quantum-limited strain sensitivity, future observing runs target higher circulating arm powers and higher squeezing levels. 
For higher circulating power, it is essential to minimize thermal transients described in \cref{sec:highpower}. An upgrade of the thermal compensation system is planned, including a new central CO$_2$ laser to reduce thermal transients from central heating. 
Increasing the circulating power may also require higher laser powers from the pre-stabilized laser system. 
Currently, two solid-state amplifiers are used for a maximum input laser power of 110 W; it is expected that 200 W of input power could be achieved by adding a second parallel amplification path of the same design and coherently combining them~\cite{CBC}. 
Further reducing quantum noise with squeezing will require addressing the limitations described in \cref{sec:sqz_characterization}; in particular, reaching the A+ goal for \qty{6}{dB} of stable and broadband squeezing will require robust and active control of misalignment and mode-mismatch, and if possible, further reducing detector-wide losses.

As part of the A+ program for quantum enhancement, a balanced homodyne readout scheme will be implemented in upcoming observing runs~\cite{Fritschel:14}.
This new readout scheme allows the homodyne angle to be chosen as a free parameter to optimize quantum noise, and mitigate several noise couplings from auxiliary length controls and laser noise. 

The O4 observing run has deepened our understanding of the universe, and will more than double the catalog of observed gravitational-wave events.
Further detector upgrades and improvements will continue to expand the astrophysical range of these detectors, contributing further to the development of gravitational-wave astronomy. Both the successes and challenges faced by the LIGO detectors during O4 will greatly inform the strategies applied to future improvements and detectors as the field of gravitational-wave astronomy is further revolutionized.

\begin{acknowledgments} 
The authors gratefully acknowledge the support of the United States National Science Foundation (NSF) for the construction and operation of the LIGO Laboratory and Advanced LIGO as well as the Science and Technology Facilities Council (STFC) of the United Kingdom, and the Max-Planck-Society (MPS) for support of the construction of Advanced LIGO. Additional support for Advanced LIGO was provided by Australian Research Council award LE210100002. The authors acknowledge the LIGO Scientific Collaboration Fellows program for additional support under award numbers PHY-1912598 and PHY-2309212. LIGO was constructed by the California Institute of Technology and Massachusetts Institute of Technology with funding from the National Science Foundation, and operates under cooperative agreement PHY-2309200. Advanced LIGO was built under award PHY-18680823459. The A+ Upgrade to Advanced LIGO is supported by US NSF award PHY-1834382 and UK STFC award ST/S00246/1, with additional support from the Australian Research Council.
This document carries LIGO Document number P2400256.
\end{acknowledgments}

\onecolumngrid

\twocolumngrid


\bibliography{references}

\end{document}